\documentclass[a4paper,11pt]{article}
\pdfoutput=1

\usepackage{jcappub}
\usepackage[T1]{fontenc}

\usepackage{physics}
\usepackage{siunitx}
\usepackage{booktabs}
\usepackage{array}
\usepackage{colortbl}

\definecolor{grey}{cmyk}{0,0,0,0.2}
\newcommand{\PreserveBackslash}[1]{\let\temp=\\#1\let\\=\temp}
\newcolumntype{C}[1]{>{\PreserveBackslash\centering}p{#1}}
\newcolumntype{R}[1]{>{\PreserveBackslash\raggedleft}p{#1}}
\newcolumntype{L}[1]{>{\PreserveBackslash\raggedright}p{#1}}

\newcommand{\cV}{\mathcal{V}}
\newcommand{\MPl}{M_\mathrm{Pl}}
\newcommand{\Pcond}[4]{P(#1,#2 \mspace{1mu} | \mspace{1mu} #3,#4)}
\newcommand{\cL}{\mathcal{L}}
\newcommand{\cLd}{\mathcal{L}^\dagger}
\newcommand{\PFPT}{P_\text{FPT}}
\newcommand{\tPFPT}{\tilde{P}_\text{FPT}}
\newcommand{\boundA}{{b1}}
\newcommand{\boundB}{{b2}}
\newcommand{\subboundA}{{\tilde{b}1}}
\newcommand{\subboundB}{{\tilde{b}2}}
\newcommand{\R}{\mathcal{R}}
\newcommand{\PR}{\mathcal{P}_\R}
\newcommand{\Pphi}{\mathcal{P}_\phi}
\newcommand{\iFi}{{}_1 F_1}
\newcommand{\oFi}{{}_0 F_1}
\newcommand{\tFt}{{}_2 F_2}
\newcommand{\exptV}{\langle\tilde{V}\rangle}

\title{Eternal inflation near inflection points: a challenge to primordial black hole models}
\author[a,b]{Eemeli Tomberg}
\author[a]{Konstantinos Dimopoulos}

\affiliation[a]{Consortium for Fundamental Physics, Physics Department,\\Lancaster University, Lancaster LA1 4YB, United Kingdom.}

\affiliation[b]{Cosmology, Universe and Relativity at Louvain (CURL), Institute of Mathematics and Physics, University of Louvain, 2 Chemin du Cyclotron, 1348 Louvain-la-Neuve, Belgium}

\emailAdd{eemeli.tomberg@uclouvain.be}
\emailAdd{k.dimopoulos1@lancaster.ac.uk}

\abstract{
Inflation with an inflection point potential is a popular model for producing primordial black holes. The potential near the inflection point is approximately flat, with a local maximum next to a local minimum, prone to eternal inflation. We show that a sufficient condition for eternal inflation is $\lambda_1 \leq 3$, where $\lambda_1$ is the index of the `exponential tail,' the lowest eigenvalue of the Fokker--Planck equation over a bounded region. We write $\lambda_1$ in terms of the model parameters for linear and quadratic regions. Wide quadratic regions inflate eternally if the second slow-roll parameter $\eta_V \geq -6$. We test example models from the literature and show this condition is satisfied; we argue eternal inflation is difficult to avoid in inflection point PBH models. Eternally inflating regions correspond to type II perturbations and form baby universes, hidden behind black hole horizons. These baby universes are inhomogeneous on large scales and dominate the multiverse's total volume. We argue that, if volume weighting is used, eternal inflation makes inflection point primordial black hole models incompatible with large-scale structure observations.}

\begin{document}

\maketitle

\section{Introduction}
\label{sec:intro}
Cosmic inflation \cite{Starobinsky:1980te,Kazanas:1980tx,Sato:1981qmu,Guth:1980zm} explains the observed large-scale homogeneity of the Universe, as well as the small inhomogeneities of quantum origin that get imprinted on the cosmic microwave background (CMB) radiation \cite{Planck:2018jri, ACT:2025fju} and grow to form the structures we see around us. On short scales, the inhomogeneities could be strong, leading to an early formation of \emph{primordial black holes} (PBHs) \cite{Carr:1974nx, Carr:1975qj}. Black hole forming inflationary models are interesting since the PBHs may, e.g, be the dark matter, seed the supermassive black holes observed in the centers of galaxies, and form binaries that emit gravitational waves \cite{Green:2024bam, Carr:2025kdk}.

The large perturbations that form PBHs can also lead to \emph{eternal inflation}. Inflation is typically driven by a scalar field called the inflaton, which rolls towards its end-of-inflation value, followed by reheating. Perturbations add a stochastic quantum noise to the classical rolling. In rare patches of space, the noise can oppose the classical drift, so that reheating is delayed. Inflating patches expand fast (faster than reheated ones), so that, if the patches are not too rare and the delay is long enough, the inflating volume can remain non-zero even at late times (in these cases, it typically grows without limit and dominates over the reheated volume). This is called eternal inflation \cite{Steinhardt:1982kg, Vilenkin:1983xq, Linde:1986fc, Linde:1986fd, Vilenkin:1999pi, Vilenkin:2004vx, Guth:2007ng, Winitzki:2008zz, Linde:2015edk}. Eternally inflating spacetimes can have a complicated fractal structure \cite{Aryal:1987vn, Winitzki:2001np, Jain:2019gsq}, and making cosmological predictions in them can be difficult due to the measure problem \cite{Linde:1994gy, Winitzki:2006rn, Freivogel:2011eg}, the difficulty of attaching probabilities to reheated patches in the infinite fractal.

In this paper, we study eternal inflation and its implications in single-field inflection point models of inflation. We consider models where, near the end of inflation,
the inflaton potential has a local minimum followed by a local maximum; as the inflaton rolls over this feature, its velocity decreases and the perturbations grow, which can lead to copious PBH production \cite{Karam:2022nym}. The perturbations' probability distribution is described by the Fokker--Planck equation. At late times, it exhibits an exponential tail \cite{Pattison:2017mbe, Ezquiaga:2019ftu, Tomberg:2023kli}, with an index $\lambda_1$ equal to the leading eigenvalue of the Fokker--Planck operator. We show rigorously that $\lambda_1 \leq 3$ is a sufficient condition for eternal inflation, also in a case where we restrict our attention only to the vicinity of the above-mentioned feature in the potential. We approximate the potential in this vicinity with linear and quadratic forms, allowing us to solve $\lambda_1$ analytically.

We then apply our solutions to example inflection point models from the literature and show that they all undergo eternal inflation. We argue this is a general result in such models, and discuss the implications for the models' CMB predictions in light of the eternally inflating fractal. It turns out that typical reheated patches are highly inhomogeneous at large scales, incompatible with the CMB observations.\footnote{The crucial word here is `typical', by which we assume volume weighting of probabilities.} This challenges the viability of inflection point PBH models and, more generally, the viability of all eternally inflating PBH models.

Throughout the paper, we assume canonical single-field inflation and use the following notations for the slow-roll parameters,
\begin{equation} \label{eq:potential_SR_parameters}
    \epsilon_V \equiv \frac{\MPl^2}{2}\qty(\frac{V'(\phi)}{V(\phi)})^2 \, , \qquad
    \eta_V \equiv  \MPl^2 \frac{V''(\phi)}{V(\phi)} \, ,
\end{equation}
where $V(\phi)$ is the potential of the inflaton field $\phi$, $\MPl$ is the reduced Planck mass, a prime denotes a derivative with respect to the argument $\phi$, and
\begin{equation} \label{eq:hubble_SR_parameters}
    \epsilon_H \equiv \frac{\dot\phi^2}{2H^2\MPl^2} = -\frac{\dot H}{H^2} \, , \qquad
    \eta_H \equiv -\frac{\ddot\phi}{H\dot\phi} \, ,
\end{equation}
where a dot denotes a derivative with respect to the cosmic time $t$ and $H$ is the Hubble parameter. In slow roll, when all the parameters are small, we have $\epsilon_V \approx \epsilon_H$, $\eta_V \approx \eta_H + \epsilon_H$. To make comparisons with the literature easier, let us also give an alternate common form of the second slow-roll parameter:
\begin{equation} \label{eq:epsilon_2}
    \epsilon_2 \equiv \frac{\dot \epsilon_H}{H \epsilon_H} = 2\qty(\epsilon_H - \eta_H) \, ,
\end{equation}
where the last equality holds beyond the slow-roll approximation.

The paper is organized as follows. In Section~\ref{sec:fokker_planck}, we develop the stochastic inflation formalism needed to quantify eternal inflation. In Section~\ref{sec:analytical_solutions}, we find our analytical solutions for simple potentials, and in Section~\ref{sec:numerical} we apply them to example PBH potentials. We discuss the consequences of eternal inflation in Section~\ref{sec:discussion} and compare our findings to previous literature in Section~\ref{sec:literature_comparison}, finally concluding in Section~\ref{sec:conclusions}.

\section{Fokker--Planck equation and eternal inflation}
\label{sec:fokker_planck}
Let us study the stochastic dynamics of the inflaton field $\phi$, using the number of e-folds of expansion $N$ as the time variable\footnote{For motivation to use $N$ as the time variable in stochastic inflation, see \cite{Finelli:2008zg, Vennin:2015hra}.} ($\dd N = H \dd t$). The field's probability distribution $P(\phi,N)$ follows the Fokker--Planck equation (see, e.g., \cite{Vennin:2020kng, Vennin:2024yzl})
\begin{equation} \label{eq:fokker_planck}
    \partial_N P(\phi,N) = \underbrace{\partial_\phi\qty[\partial_\phi\qty(\frac{1}{2} \sigma^2(\phi) P(\phi,N)) + \cV'(\phi) P(\phi,N)]}_{\equiv \cL_{\text{FP},\phi}P(\phi,N)} \, ,
\end{equation}
where we introduced the Fokker--Planck operator $\cL_{\text{FP},\phi}$. Here $\cV$ is the `effective potential' (not to be confused with the inflaton potential $V$) that gives the deterministic drift through $\partial_N\phi = -\cV'(\phi)$ when stochastic effects are disregarded, and $\sigma$ is the diffusion strength that controls the field's variance through $\partial_N \expval{\phi^2} = \expval{\sigma^2(\phi)}$ when the drift is disregarded.
Equation \eqref{eq:fokker_planck} applies if the field evolution is Markovian, that is, if the stochastic noise only depends on the current field value $\phi$. This is true, in particular, during slow-roll inflation, where
\begin{equation} \label{eq:SR_sigma_V}
    \sigma(\phi) = \frac{H(\phi)}{2\pi} = \frac{\sqrt{V(\phi)}}{2\sqrt{3}\pi\MPl} \, , \qquad \cV'(\phi) = \frac{V'(\phi)\MPl^2}{V(\phi)} \, .
\end{equation}
See, e.g., \cite{Tomberg:2024evi} for deriving the Fokker--Planck equation from first principles. Below, we will go beyond slow roll by considering a constant-roll approximation, which modifies \eqref{eq:SR_sigma_V}; for now, we stick with the generic form \eqref{eq:fokker_planck}.

To solve the Fokker--Planck equation \eqref{eq:fokker_planck}, we need to set boundary conditions. In the applications below, when a stochastic path hits a boundary $\phi_b$, we want to remove it from the computation. This can be achieved with an absorbing boundary: $P(\phi_{b},N)=0$ for all $N$.\footnote{Physically, this corresponds to increasing the drift $-\cV'$ to (plus or minus) infinity beyond the boundary, so that paths venturing there are immediately swept away at an infinite velocity, regardless of the diffusion. When the velocity increases, $P(\phi, N)$ must decrease to zero to keep the flux finite.} For stochastic motion between two absorbing boundaries at $\phi_\boundA$ and $\phi_\boundB$ (one of which may be taken to $\pm\infty$), we define the survival probability
\begin{equation} \label{eq:survival_probability}
    S(N)
    = \int_{\phi_{\phi_\boundA}}^{\phi_\boundB} P(N,\phi) \dd \phi \, ,
\end{equation}
which is the fraction of spatial patches still within the interval at time $N$. As $N$ increases, more and more patches exit the interval, and $S(N)$ approaches zero.

In the next sections (until \ref{sec:sub_potential}), we take $\phi_\boundA$ and $\phi_\boundB$ to mark end-of-inflation surfaces: once inflation stops, stochastic motion ceases, and the field can't return to the inflating region. Then $S(N)$ gives the fraction of space still inflating at $N$, useful for considerations about eternal inflation.

\subsection{Eternal inflation}
\label{sec:eternal_inflation}
Eternal inflation means that some portion of the Universe undergoes inflation at any given time, even in the far future. However, it is not clear what `some portion of the Universe' or a `given time' mean in this statement. To make progress, one has to choose a foliation of space-time (to fix a time coordinate) and a measure of spatial regions within this foliation.\footnote{The arbitrariness of these choices has been discussed in, e.g., \cite{Linde:1993xx, Garcia-Bellido:1994gng, Winitzki:2001np, Winitzki:2005ya}, and it is related to the measure problem, which we will return to in Section~\ref{sec:discussion}.}

We choose a foliation with the number of e-folds $N$ as the time coordinate. Starting from a finite inflating patch, we say that eternal inflation happens if the expected physical volume of inflating space within the patch does not vanish as $N \to \infty$ (see, e.g., \cite{Linde:1993xx,  Linde:1993nz, Winitzki:2001np, Rudelius:2019cfh} for similar takes). The volume grows with $N$ as $e^{3N}$, and the expectation value can be computed as an integral over $P(\phi,N)$, so we get
\begin{equation} \label{eq:eternal_inflation_condition_1}
    \text{eternal inflation} \quad \iff \quad
    \lim_{N\to\infty} \expval{V}_N \equiv \lim_{N\to\infty} \int_{\phi_\boundA}^{\phi_\boundB} e^{3N} P(\phi,N) \dd \phi > 0 \, .
\end{equation}
Note that $e^{3N}$ does not depend on $\phi$, so the integral is proportional to the survival probability $S(N)$, the fraction of space still inflating; eternal inflation takes place if $S(N)$ -- and thus, $P(\phi,N)$ -- does not decrease faster than $e^{-3N}$ in the large-$N$ limit.

It was shown in \cite{Winitzki:2001np, Winitzki:2005ya} that foliations with other time coordinates produce equivalent conclusions about eternal inflation, as long as the change of coordinates depends only on the field $\phi$. In a more general foliation, the situation becomes more complicated. In Appendix~\ref{sec:first_passage_times}, we introduce an alternative way to describe eternal inflation, based on the expected volume of the reheating surface, and thus free of such complications \cite{Creminelli:2008es, Tegmark:2004qd}. In Appendix~\ref{sec:sturm-liouville}, we show this description is equivalent to \eqref{eq:eternal_inflation_condition_1}.

\subsection{Late-time behaviour}
\label{sec:late_time_behaviour}
To solve the Fokker--Planck equation \eqref{eq:fokker_planck}, we seek the spectrum of the Fokker--Planck operator $\cL_{\text{FP},\phi}$, that is, eigenvalues $\lambda$ and the corresponding eigenfunctions $u(\phi)$ that satisfy\footnote{The minus sign in \eqref{eq:FP_eigenfunctions} is a typical convention for Sturm--Liouville problems.}
\begin{equation} \label{eq:FP_eigenfunctions}
    \cL_{\text{FP},\phi} u(\phi) = -\lambda u(\phi) \, .
\end{equation}
The eigenvalues are described by \emph{Strum--Liouville theory}. With absorbing boundary conditions, they are all positive. The eigenenvectors  $u_n(\phi)$ form a complete, orthogonal basis, and let us write $P(\phi,N)$ as
\begin{equation} \label{eq:P_expanded_in_u}
    P(\phi,N) = \sum_n a_n(N) u_n(\phi) \, ,
\end{equation}
so that, by the Fokker--Planck equation \eqref{eq:fokker_planck},
\begin{equation} \label{eq:P_expanded_in_u_evolution}
\begin{aligned}
    &\partial_N P(\phi,N)
    = \sum_n a_n'(N) u_n(\phi) \\
    &\phantom{\partial_N P(\phi,N)} = \sum_n a_n(N) \cL_{\text{FP},\phi} u_n(\phi)
    = \sum_n -\lambda_n a_n(N) u_n(\phi) \\
    &\implies
    a_n(N) = c_n e^{-\lambda_n N} \, ,
\end{aligned}
\end{equation}
where $c_n$ are constants that depend on the initial conditions. We discuss the details of this construction in Appendix~\ref{sec:sturm-liouville}.

At late times, the lowest eigenvalue $\lambda_1$ dominates over the others, so $P(\phi,N) \to c_1 u_1(\phi) e^{-\lambda_1 N}$.
The eternal inflation condition \eqref{eq:eternal_inflation_condition_1} then becomes
\begin{equation} \label{eq:eternal_inflation_condition_in_lambda}
    \setlength{\fboxsep}{5\fboxsep}\boxed{\text{eternal inflation} \quad \iff \quad \lambda_1 \leq 3 \, .}
\end{equation}
This is not sensitive to the initial conditions; it only depends on the inflaton potential.

\subsection{Studying a subsection of the potential}
\label{sec:sub_potential}

Many inflationary potentials support eternal inflation at extreme field values \cite{Greenwood:2021uuj}. Indeed, the CMB observations reveal a red-tilted power spectrum \cite{Planck:2018jri, ACT:2025fju}, implying the primordial perturbations grow at long distances; if there is no stop to this growth, one eventually hits an eternally inflating regime. Instead of such long-distance properties of a model, we are interested in eternal inflation close to the potential's PBH-forming feature. We also note that solving the minimal eigenvalue $\lambda_1$ is not easy for a full, general potential: there may be many model parameters to scan over, and a numerical solution may be difficult if the feature is much narrower than the full field range, as we will discuss in more detail in Section~\ref{sec:numerical}. However, it may be possible to find a solution for a restricted subsection of the field space around the feature, if the potential can be approximated there by some simple function.

For these reasons, let us next study stochastic dynamics restricted to field range $\phi \in [\phi_\subboundA,\phi_\subboundB]$, a subsection of the full inflating range $[\phi_\boundA,\phi_\boundB]$. We solve the Fokker--Planck equation in this subsection, with absorbing boundary conditions at both ends, and call the lowest eigenvalue $\tilde{\lambda}_1$. If $\tilde{\lambda}_1 \leq 3$, we have eternal inflation in the subsection.\footnote{Concentrating on sub-potentials in eternal inflation was brought up in \cite{Creminelli:2008es}; here we develop the idea further.}

Physically, when solving for $\tilde{\lambda}_1$, we're restricting our attention to stochastic paths that stay in the interval $[\phi_\boundA,\phi_\boundB]$ for a long time. Consider the survival probabilities
\begin{equation} \label{eq:survival_probabilities}
\begin{aligned}
    S(N)
    = \int_{\phi_\boundA}^{\phi_\boundB} P(N,\phi) \dd \phi \sim e^{-\lambda_1 N}  \, ,
    \qquad
    \tilde{S}(N)
    = \int_{\phi_\subboundA}^{\phi_\subboundB} \tilde{P}(N,\phi) \dd \phi \sim e^{-\tilde{\lambda}_1 N} \, ,
\end{aligned}
\end{equation}
where $P$ is computed with the $[\phi_\boundA,\phi_\boundB]$ boundaries and $\tilde{P}$ with the $[\phi_\subboundA,\phi_\subboundB]$ boundaries, and we have indicated the large-$N$ behaviours. Starting with the initial field within $[\phi_\subboundA,\phi_\subboundB]$, we must have
\begin{equation} \label{eq:survival_probability_hierarchy}
    S(N) > \tilde{S}(N) \quad \text{for any $N$} \quad \implies \quad \lambda_1 \leq \tilde{\lambda}_1 \, ,
\end{equation}
since each path must first pass out of the inner boundaries $[\phi_\subboundA,\phi_\subboundB]$ before passing out of the outer boundaries $[\phi_\boundA,\phi_\boundB]$. Using \eqref{eq:survival_probability_hierarchy}, we see that $\tilde{\lambda}_1 \leq 3$ is a \emph{sufficient} condition for eternal inflation in the full potential, as per \eqref{eq:eternal_inflation_condition_in_lambda}.
This eternal inflation happens within the restricted field range $[\phi_\subboundA,\phi_\subboundB]$: the expected volume of all paths that stay within this range for a long time stays non-zero as $N \to \infty$. This obviously guarantees the same is true for the full set of paths, yielding eternal inflation as defined in \eqref{eq:eternal_inflation_condition_1}. The logic does not depend on the exact long-time behaviour; for completeness, in Appendix~\ref{sec:eternal_inflation_from_subpotential} we prove that eternal inflation in a sub-potential guarantees eternal inflation in the full potential without assuming $S(N)\propto e^{-\lambda_1 N}$  and with initial conditions outside $[\phi_\subboundA,\phi_\subboundB]$.

Let us next consider a number of different `sub-potentials' with restricted field ranges and solve $\lambda_1$ for them (for simplicity, we drop the tildes from the subsection quantities from now on). We will use these results to give sufficient conditions for eternal inflation in typical single-field PBH models.

\section{Analytical solutions for sub-potentials}
\label{sec:analytical_solutions}

\subsection{Constant drift}
\label{sec:constant_drift}
Let us start with a simple case with
\begin{equation} \label{eq:constant_drift_diffusion_drift}
    \sigma = \text{const.} \, , \qquad \cV' = \alpha = \text{const.}
\end{equation}
Let us set absorbing boundaries at 0 and $2\phi_b$ (due to shift symmetry, this is equal to placing symmetric boundaries at $\pm\phi_b$). Now, the Fokker-Planck eigenvalue equation $\cL_{\text{FP},\phi}u(\phi) = -\lambda u(\phi)$ has the general solution
\begin{equation} \label{eq:const_drift_u}
    u(\phi) = e^{-\frac{\alpha\phi}{\sigma^2}}\qty[c_1 \sin(\frac{\sqrt{2\sigma^2\lambda-\alpha^2}\phi}{\sigma^2})
    + c_2 \cos(\frac{\sqrt{2\sigma^2\lambda-\alpha^2}\phi}{\sigma^2})] \, .
\end{equation}
The first boundary condition $u(0)=0$ sets the integration constant $c_2 = 0$. The second boundary condition sets
\begin{equation} \label{eq:const_drift_lambda}
    \frac{\sqrt{2\sigma^2\lambda_n-\alpha^2}}{\sigma^2}\times 2\phi_b = \pi n \implies
    \lambda_n = n^2\pi^2\frac{\sigma^2}{8\phi_b^2} + \frac{\alpha^2}{2\sigma^2} \, , \quad n=1,2,\dots
\end{equation}
where we indeed found a numerable set of eigenvalues $\lambda_n$. A similar result was obtained in, e.g., \cite{Ezquiaga:2019ftu}. The first term in \eqref{eq:const_drift_lambda} describes the effects of the finite width $2\phi_b$. When $2\phi_b$ is taken to infinity, we find the absolute minimum for the lowest eigenvalue, the second term, $\lambda_0 \equiv \alpha^2/(2\sigma^2)$.\footnote{Note that $\lambda > \alpha^2/(2\sigma^2)$ is required, since otherwise, the trigonometric functions in \eqref{eq:const_drift_u} turn into hyperbolic ones, which don't have the required two zeros for real arguments.} As $\phi_b \to \infty$, all the eigenvalues approach $\lambda_0$, transforming the discrete spectrum into a continuous one (in this limit, the Sturm--Liouville problem is no longer regular, see Appendix~\ref{sec:sturm-liouville}).

\paragraph{Slow roll.}
The constant drift \eqref{eq:constant_drift_diffusion_drift} corresponds to slow-roll inflation with the potential
\begin{equation} \label{eq:constant_drift_V}
    V=V_0\qty(1+\frac{\sqrt{2\epsilon_V}\phi}{\MPl})
\end{equation}
for constant $V_0$ and $\epsilon_V$ (the potential and first potential slow-roll parameter at $\phi=0$).\footnote{In the context of constant drift, we use $\epsilon_V$ to refer to the value at the origin, even though the first slow-roll parameter slightly varies around this point.} Matching with \eqref{eq:constant_drift_diffusion_drift} and \eqref{eq:SR_sigma_V}, we get $\epsilon_V = \alpha^2/(2\MPl^2)$.
We demand $\sqrt{2\epsilon_V}\phi/\MPl \ll 1$ for $0<\phi<2\phi_b$, so that the height of the potential is approximately constant for the purposes of computing the diffusion coefficient $\sigma=H/(2\pi)=\sqrt{V_0}/(2\sqrt{3}\pi\MPl)$; this sets $\phi_b \ll \MPl/\sqrt{8\epsilon_V}$. We also demand $\epsilon_V \ll 1$ so that the slow-roll approximation applies.

The leading eigenvalue becomes
\begin{equation} \label{eq:lambda_1_constant_drift}
    \lambda_1 = \frac{H^2}{32 \phi_b^2} + \frac{4\pi^2 \MPl^2 \epsilon_V}{H^2} \, .
\end{equation}
In the slow-roll approximation, the curvature power spectrum $\PR = H^2/(8\pi^2\MPl^2\epsilon_V) = \sigma^2/\alpha^2 = 1/(2\lambda_0)$. In the wide field-range limit, $\lambda_1 = \lambda_0 \leq 3$ then gives:
\begin{equation} \label{eq:eternal_inflation_condition_linear_sr}
    \setlength{\fboxsep}{5\fboxsep}\boxed{
    \phi_b \gg H \, , \quad
    \epsilon_V \ll 1 \, , \quad
    \PR \geq 1/6 \, \quad \implies \quad \text{eternal inflation in a linear potential}\, .}
\end{equation}
This agrees with the heuristic, approximate condition $\PR \gtrsim 1$ for eternal inflation \cite{Linde:2015edk, Creminelli:2008es}. When $\phi_b$ becomes small, the true lowest eigenvalue $\lambda_1$ becomes larger than $\lambda_0$, suppressing eternal inflation even for large $\PR$.

\subsection{Linear drift}
\label{sec:linear_drift}
Let us next consider the case with a linear drift,
\begin{equation} \label{eq:linear_drift_diffusion_drift}
    \sigma = \text{const.} \, , \qquad \cV' = \beta\phi \, .
\end{equation}
Let us also set symmetric absorbing boundary conditions at $\pm\phi_b$. We will first discuss the general problem in terms of the constants $\sigma$, $\beta$, and $\pm\phi_b$; later, we will connect these to inflationary potentials, where $\beta$ is connected to the second slow-roll parameter, so that $\beta < 0$ corresponds to a quadratic maximum and $\beta > 0$ to a quadratic minimum.

For convenience, instead of the Fokker--Planck operator $\cL_{\text{FP},\phi}$, we work with its adjoint, finding the solutions of $\cLd_{\text{FP},\phi}\bar{u}(\phi) = -\lambda \bar{u}(\phi)$ (see Appendix~\ref{sec:first_passage_times} and equation~\eqref{eq:L_FP}). As adjoints, $\cL_{\text{FP},\phi}$ and $\cLd_{\text{FP},\phi}$ share eigenvalues. With the change of variables $\varphi = \beta\phi^2/\sigma^2$, the $\cLd_{\text{FP},\phi}$ eigenvalue equation takes the form of \emph{Kummer's equation}
\begin{equation} \label{eq:kummer}
    \varphi \partial^2_\varphi \bar{u}(\varphi) + \qty(\frac{1}{2} - \varphi)\partial_\varphi \bar{u} + \frac{\lambda}{2\beta}\bar{u}(\varphi) = 0 \, ,
\end{equation}
solved by confluent hypergeometric functions $\iFi(a;b;z)$. The full solution is
\begin{equation} \label{eq:kummer_solved}
    \bar{u}(\phi) = c_1 \, \iFi\qty(-\frac{\lambda}{2\beta}; \frac{1}{2}; \frac{\phi^2}{\sigma^2}\beta)
    + c_2 \, \phi \, \iFi\qty(-\frac{\lambda}{2\beta}+\frac{1}{2}; \frac{3}{2}; \frac{\phi^2}{\sigma^2}\beta) \, .
\end{equation}
The first branch is even in $\phi$, and the second branch is odd. The boundary conditions $\bar{u}(\pm\phi_b)=0$ dictate that both branches must vanish individually, by setting either the $c_i$ coefficient or the corresponding $\iFi$ function to zero. In practice then, the two branches produce independent eigensolutions with independent $\lambda$ values. Sturm--Liouville theory tells us that the lowest solution should have no zeroes in the interval $]-\phi_b,\phi_b[$ (see Appendix~\ref{sec:sturm-liouville}), in other words, it must come from the even branch instead of the odd one. (The number of zeroes increases by one for each subsequent solution, so the solutions alternate between the branches.) We thus seek the lowest $\lambda$ for which
\begin{equation} \label{eq:linear_drift_eigenvalues}
    \iFi\qty(-\frac{\lambda}{2\beta}; \frac{1}{2}; \frac{\phi_b^2}{\sigma^2}\beta) = 0 \, .
\end{equation}
The confluent hypergeometric functions have the series representation
\begin{equation} \label{eq:hypergeometric_series}
    \iFi\qty(a; b; z) = \sum_{n=0}^{\infty} \frac{a^{(n)} z^n}{b^{(n)}n!} \, ,
\end{equation}
where the rising factorial $x^{(n)}$ is defined as
\begin{equation} \label{eq:rising_factorial}
    x^{(0)} \equiv 1 \, , \quad x^{(n)} \equiv x(x+1)(x+2)\dots(x+n-1) \;\, \text{for a positive integer $n$.}
\end{equation}
The functions obey \emph{Kummer's transformation},\footnote{In our context, Kummer's transformation is related to transforming between eigenfunctions of the adjoint and usual Fokker--Planck equations, see \eqref{eq:choosing_p_q_w} and \eqref{eq:FP_vs_adjoint_eigenstates}, as the exponential factor suggests.}
\begin{equation} \label{eq:kummers_transformation}
    \iFi(a; b; z) = e^z \iFi(b-a; b; -z) \, ,
\end{equation}
so \eqref{eq:linear_drift_eigenvalues} can be written in the equivalent form
\begin{equation} \label{eq:linear_drift_eigenvalues_2}
    \iFi\qty(\frac{1}{2}+\frac{\lambda}{2\beta}; \frac{1}{2}; -\frac{\phi_b^2}{\sigma^2}\beta) = 0 \, .
\end{equation}

\paragraph{Negative $\boldsymbol\beta$.}
The solution of \eqref{eq:linear_drift_eigenvalues} and \eqref{eq:linear_drift_eigenvalues_2} depends on the sign of $\beta$. Let us first consider the case $\beta<0$, that is, repulsive drift away from $\phi=0$, corresponding to a hilltop potential. If also $\lambda \leq 0$, then all the arguments in \eqref{eq:linear_drift_eigenvalues_2} are positive, as clearly is the sum \eqref{eq:hypergeometric_series}. Thus, solutions only exist for $\lambda>0$, in accordance with the absorbing boundaries -- to be more precise, for $\lambda > -\beta$. For these values, the first argument in \eqref{eq:linear_drift_eigenvalues_2} is negative, so different terms in the series \eqref{eq:hypergeometric_series} have different signs, and the sum may yield zero for many discrete values of $\lambda$.

\begin{figure}
    \centering
    \includegraphics[]{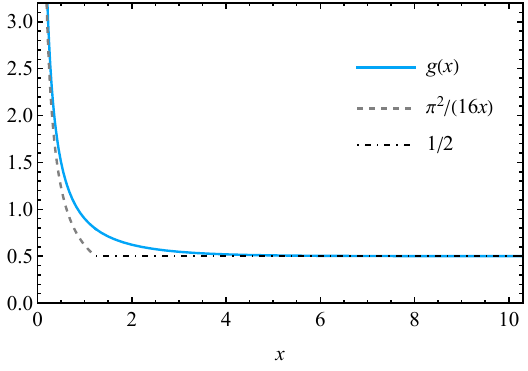}
    \caption{The function $g$ defined in \eqref{eq:g_defined}, with the asymptotic limits \eqref{eq:g_asymptotics}.}
    \label{fig:g}
\end{figure}

Instead of \eqref{eq:linear_drift_eigenvalues_2}, let us turn our attention to the form \eqref{eq:linear_drift_eigenvalues} and write it schematically as
\begin{equation} \label{eq:g_defined}
    \iFi\qty(g(x); \frac{1}{2}; -x) = 0
\end{equation}
where $g(x)$ is the lowest solution which satisfies the equation for a given $x>0$, and the lowest eigenvalue takes the form
\begin{equation} \label{eq:lambda_in_g_beta_negative}
    \lambda_1 = 2|\beta| g\qty(\frac{\phi_b^2}{\sigma^2}|\beta|) \quad \text{for} \quad \beta < 0 \, .
\end{equation}
We argued above that $\lambda_1 > 0$, so $g(x)>0$, that is, $g$ is a function from non-negative real numbers to non-negative real numbers. It can be solved numerically from \eqref{eq:g_defined}; the solution it is depicted in Figure~\ref{fig:g}.

In Appendix~\ref{sec:g_properties} we show that, asymptotically,
\begin{equation} \label{eq:g_asymptotics}
    g(x) \xrightarrow{x \to 0^+} \frac{\pi^2}{16 x} \, , \qquad
    g(x) \xrightarrow{x \to \infty} \frac{1}{2} \, .
\end{equation}
The second limit is approached roughly exponentially. For $\lambda_1$, these imply
\begin{equation} \label{eq:lambda_1_asymptotics_beta_negative}
    \lambda_1 \xrightarrow{\frac{\phi_b^2}{\sigma^2}|\beta| \to 0^+} \frac{\pi^2\sigma^2}{8\phi_b^2} \, , \quad
    \lambda_1 \xrightarrow{\frac{\phi_b^2}{\sigma^2}|\beta| \to \infty} |\beta| \quad \text{for} \quad \beta < 0 \, .
\end{equation}
The first limit in \eqref{eq:lambda_1_asymptotics_beta_negative} diverges in a way that does not depend on $\beta$. It corresponds to a flat potential: either $|\beta| \ll 1$ so that the potential's curvature vanishes, or $\phi_b \ll \sigma$ so that the field range under consideration is much narrower than the typical field variations given by the diffusion strength $\sigma$. This agrees with the flat potential $\alpha \to 0$ limit of constant drift, see \eqref{eq:const_drift_lambda}. The second limit in \eqref{eq:lambda_1_asymptotics_beta_negative} does not depend on $\phi_b$ or $\sigma$. It corresponds either to $|\beta| \gg 1$ or $\phi_b \gg \sigma$; in either case, the field gets to explore essentially the full curve of the potential, not feeling the boundaries.

\paragraph{Positive $\boldsymbol\beta$.}
Let us then consider the case $\beta > 0$, that is, attractive drift towards $\phi=0$, the bottom of a potential well. Our argument proceeds as in the $\beta < 0$ case, but with the roles of the two forms \eqref{eq:linear_drift_eigenvalues} and \eqref{eq:linear_drift_eigenvalues_2} reversed. Now the form \eqref{eq:linear_drift_eigenvalues} has all positive arguments for $\lambda < 0$, showing that solutions can only exist for $\lambda \geq 0$. For $\lambda>0$, \eqref{eq:linear_drift_eigenvalues_2} has a form analogous to \eqref{eq:g_defined} (that is, a positive first argument and a negative last argument), and we get
\begin{equation} \label{eq:lambda_in_g_beta_positive}
    \lambda_1 = \beta \qty[2g\qty(\frac{\phi_b^2}{\sigma^2}\beta) - 1] \quad \text{for} \quad \beta > 0
\end{equation}
with $g(x)$ the same function as above. This time, the limits \eqref{eq:g_asymptotics} give
\begin{equation} \label{eq:lambda_1_asymptotics_beta_positive}
    \lambda_1 \xrightarrow{\frac{\phi_b^2}{\sigma^2}\beta \to 0^+} \frac{\pi^2\sigma^2}{8\phi_b^2} - \beta \, , \quad
    \lambda_1 \xrightarrow{\frac{\phi_b^2}{\sigma^2}\beta \to \infty} 0 \quad \text{for} \quad \beta > 0 \, .
\end{equation}
The first limit again diverges and agrees with the flat case for $\beta \to 0$. The second limit again corresponds to the field exploring the full potential; this time, the probability distribution settles to a static equilibrium around the potential minimum and evolves no more, so $\lambda_1 = 0$. For a finite $\frac{\phi_b^2}{\sigma^2}\beta$, $\lambda_1$ is always somewhat above zero: some probability continuously leaks out of the bounds, but this process may be arbitrarily slow.

\paragraph{Example distribution.}
To demonstrate the late-time behaviour of the probability distribution, let us consider the following example:
\begin{equation} \label{eq:example_p}
    P(\phi,N) = \frac{1}{\sqrt{\pi[e^{-2\beta N} + \text{sgn}(\beta)]\sigma^2/|\beta|}}\exp(-\frac{|\beta|\phi^2}{[e^{-2\beta N} + \text{sgn}(\beta)]\sigma^2}) \, .
\end{equation}
This satisfies the Fokker--Planck equation \eqref{eq:fokker_planck} in the wide limit where $\phi_b \to \infty$. For $\beta < 0$, the distribution starts sharply peaked around $\phi=0$ at $N=0$, and spreads from there, approaching
\begin{equation} \label{eq:example_p_asymptotics_1}
    P(\phi,N) \xrightarrow{N \to \infty}
    \frac{1}{\sqrt{\pi\sigma^2/|\beta|}}\exp(-\frac{|\beta|\phi^2}{\sigma^2}e^{-2|\beta|N} - |\beta|N) \, .
\end{equation}
From here, we can read off both the leading eigenfunction $u_1(\phi)$ (a constant) and the leading eigenvalue $\lambda_1 = |\beta|$, which matches \eqref{eq:lambda_1_asymptotics_beta_negative}. For the leading behaviour to be reached, the `exponent in the exponent' must decay away; in other words, we need $N \gg 1/(2|\beta|)$.
For $\beta > 0$, the distribution starts with an infinite width at $N \to -\infty$; at late times, it settles to the equilibrium distribution
\begin{equation} \label{eq:example_p_asymptotics_2}
    P(\phi,N) \xrightarrow{N \to \infty}
    \frac{1}{\sqrt{\pi\sigma^2/\beta}}\exp(-\frac{\beta\phi^2}{\sigma^2}) \, ,
\end{equation}
giving the lowest (Gaussian) eigenfunction $u_1(\phi)$. The corresponding eigenvalue, $\lambda_1 = 0$, again matches \eqref{eq:lambda_1_asymptotics_beta_positive}. The approach is again controlled by $\beta$: the equilibrium is reached when $N \gg 1/(2\beta)$.

In Appendix~\ref{sec:g_properties}, we discuss the other values of the function $g(x)$ which satisfy equation \eqref{eq:g_defined}; these can be used to compute the second eigenvalue $\lambda_2$. In the wide limit, for either sign of $\beta$, we find $\lambda_2 - \lambda_1 = 2|\beta|$. This reinforces the intuition from the previous paragraph: the second mode has decayed away compared to the first when $(\lambda_2 - \lambda_1)N \gg 1$, that is, when $N\gg1/(2|\beta|)$.

\begin{figure}
    \centering
    \includegraphics[]{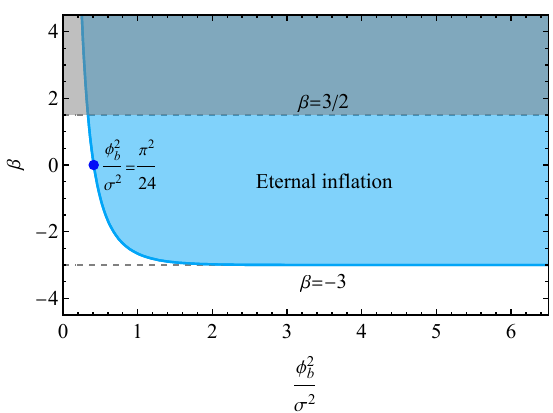}
    \caption{Parameter region for eternal inflation, $\lambda_1 \leq 3$, in the case of linear drift.}
    \label{fig:linear_drift_eternal_inflation}
\end{figure}

\paragraph{Eternal inflation.}
Let us now turn our attention to eternal inflation, which happens for $\lambda_1 \leq 3$, as per \eqref{eq:eternal_inflation_condition_in_lambda}. From \eqref{eq:lambda_in_g_beta_negative}, \eqref{eq:lambda_in_g_beta_positive}, this is equivalent to
\begin{equation} \label{eq:linear_drift_eternal_inflation}
    -\frac{3}{2g\qty(\frac{\phi_b^2}{\sigma^2}|\beta|)} < \beta < \frac{3}{2g\qty(\frac{\phi_b^2}{\sigma^2}|\beta|) - 1} \quad \text{(eternal inflation)} \, .
\end{equation}
The region is depicted in Figure~\ref{fig:linear_drift_eternal_inflation} in the $(\frac{\phi_b^2}{\sigma^2}, \beta)$ plane.
When $\frac{\phi_b^2}{\sigma^2} \ll 1$, the eternally inflating region is limited to very large $\beta$. At the boundary, to compensate the large $\beta$, $g$ approaches $1/2$ in accordance with the second limit in \eqref{eq:g_asymptotics}, so that $\lambda_1$ from \eqref{eq:lambda_in_g_beta_positive} stays small. In practice, the boundary's behaviour has to be solved numerically.
At $\frac{\phi_b^2}{\sigma^2} = \frac{\pi^2}{24}$, the boundary of the eternal inflation region crosses $\beta = 0$ -- this is the flat-potential limit, see first limits in \eqref{eq:lambda_1_asymptotics_beta_negative}, \eqref{eq:lambda_1_asymptotics_beta_positive}. For $\frac{\phi_b^2}{\sigma^2} \gg 1$, eternal inflation happens for all $\beta > -3$, in accordance with the second limit in \eqref{eq:lambda_1_asymptotics_beta_negative}.

Let us next connect $\beta$ and $\sigma$ to an inflationary potential to see if eternal inflation is expected in typical models.

\paragraph{Small $\boldsymbol\beta$: slow roll.}
Let us first consider slow-roll inflation in the following potential:
\begin{equation} \label{eq:linear_drift_V}
    V(\phi)=V_0\qty(1+\frac{\eta_V\phi^2}{2\MPl^2})
\end{equation}
for constant $V_0$ and $\eta_V$ (the potential and second potential slow-roll parameter at $\phi=0$).\footnote{In the context of linear drift, we use $\eta_V$ to refer to the value at the origin, even though the second slow-roll parameter slightly varies around this point.} This is a local quadratic minimum for $\eta_V > 0$ and a hilltop for $\eta_V < 0$. Analogously to the constant drift case, we require
\begin{equation} \label{eq:phib_requirement}
    \phi_b \ll \sqrt{\frac{2\MPl}{|\eta_V|}} \quad \implies \quad \frac{\Delta V}{V_0} = \frac{|\eta_V| \phi^2}{2\MPl^2} \ll 1 \quad \text{for all $|\phi|$ < $\phi_b$} \, ,
\end{equation}
ensuring that the potential changes only little in the region under consideration, leading to an approximately constant Hubble parameter. The slow-roll diffusion and drift \eqref{eq:SR_sigma_V} are then of the linear form \eqref{eq:linear_drift_diffusion_drift}, with the parameter values
\eqref{eq:SR_sigma_V}
\begin{equation} \label{eq:linear_drift_diffusion_drift_sr}
    \sigma = \frac{H}{2\pi} = \frac{\sqrt{V_0}}{2\sqrt{3}\pi \MPl} \, , \qquad \beta = \eta_V \, .
\end{equation}

For the slow-roll approximation to apply, we require $|\eta_V| \ll 1$ and $\epsilon_V \approx \eta_V^2\phi^2/(2\MPl^2) \ll 1$. The second condition follows automatically from the first and \eqref{eq:phib_requirement}, and \eqref{eq:phib_requirement} follows from $|\eta_V| \ll 1$ unless field excursions are vastly super-Planckian. In practice, when approximating a full potential with the quadratic form \eqref{eq:linear_drift_V}, we want to choose $\phi_b$ to be as large as possible, while respecting \eqref{eq:phib_requirement} and ensuring the parabolic approximation is good for the whole region. This maximizes the chances to find an eternally inflating region, see Figure~\ref{fig:linear_drift_eternal_inflation}. Combined with $|\beta| = |\eta_V| \ll 1$, the figure shows that eternal inflation happens for $\phi_b^2/\sigma^2 > \frac{\pi^2}{24}$, that is, for $\phi_b > H/(4\sqrt{6})$.
For single-field models where the CMB is generated in slow roll, combining the observed strength of scalar perturbations $A_s = 2.1\times10^{-9}$ \cite{Planck:2018jri} with the tensor-to-scalar ratio limit $r<0.036$ \cite{BICEP:2021xfz} gives the limit $H = \pi\sqrt{A_s r/2} \MPl \lesssim 2\times10^{-5}\MPl$, so eternal inflation is achievable even for quite small $\phi_b$. Indeed, the field is typically measured in units of $\MPl$ during inflation, so we generically expect $\phi_b \sim \MPl \gg \sigma \sim H$.

\paragraph{Large $\boldsymbol\beta$: constant roll.}
Starting from the potential \eqref{eq:linear_drift_V}, it is actually possible to relax the slow-roll condition $|\eta_V| \ll 1$. This leads to \emph{constant roll} (CR) near $\phi=0$, a phase where the second slow-roll parameter $\eta_H$ is a constant (but can be large), assuming that the condition \eqref{eq:phib_requirement} still holds, so that $H$ is approximately a constant and $\epsilon_H \ll 1$. Constant roll was originally considered in \cite{Motohashi:2014ppa}. In our setup, the constant $\eta_H$ is related to the potential as \cite{Karam:2022nym}
\begin{equation} \label{eq:eta_H_in_eta_V_CR}
    \eta_H = \frac{3}{2}\qty(1 - \sqrt{1-\frac{4}{3}\eta_V}) \, .
\end{equation}
In this setup, the diffusion and drift \eqref{eq:SR_sigma_V} still take the form \eqref{eq:linear_drift_diffusion_drift}, but with the corrected parameter values
\begin{equation} \label{eq:linear_drift_diffusion_drift_cr}
    \sigma = \frac{H}{2\pi} \times \frac{\Gamma\qty(\frac{3}{2}-\eta_H)}{\Gamma\qty(\frac{3}{2})} \times \qty(\frac{2}{\sigma_c})^{-\eta_H}
    \, , \qquad \beta = \eta_H \, .
\end{equation}
Here $\sigma_c < 1$ is a (constant) coarse-graining parameter (not to be confused with the drift $\sigma$), giving the physical size of the stochastically evolving super-Hubble patch of space by $\sim (\sigma_c H)^{-1}$. We derive \eqref{eq:linear_drift_diffusion_drift_cr} in Appendix~\ref{sec:drift_and_diffusion_in_cr} starting from the previous works \cite{Karam:2022nym, Tomberg:2023kli}. Note that when $|\eta_V|$ is small, \eqref{eq:linear_drift_diffusion_drift_cr} reverts to \eqref{eq:linear_drift_diffusion_drift_sr}.

Since $\beta$ can now be large, we can explore the full parameter space in Figure~\ref{fig:linear_drift_eternal_inflation}. Eternal inflation always requires $\beta=\eta_H \geq -3$, or $\eta_V \geq -6$, but is generic there if $\phi_b/\sigma \gg 1$.
By the same arguments as for the slow-roll case, $\phi_b/\sigma \gg 1$ in typical models, and the requirement $\eta_V \geq -6$ only discards the sharpest of hilltops.
If $\phi_b/\sigma \lesssim 1$, the parameter region is more limited and eternal inflation has to be checked case by case. One should be particularly careful when $\sigma_c \ll 1$ or $|\eta_H| \gg 1$, since $\sigma$ can grow larger than $H$ there, possibly driving $\phi_b/\sigma$ to these small values. Barring such special cases, eternal inflation seems to be a generic outcome in PBH potentials. Our most important result is then:
\begin{equation} \label{eq:eternal_inflation_condition_cr}
    \setlength{\fboxsep}{5\fboxsep}\boxed{
    \phi_b \gg H \, , \quad
    \eta_V \geq -6 \, \quad \implies \quad \text{eternal inflation in a quadratic potential}\, .}
\end{equation}

Let us comment in more detail the above-mentioned $\eta_H \to 3/2$ limit.
Note that in \eqref{eq:eta_H_in_eta_V_CR}, we need $\eta_V \leq 3/4$ or the expression gives an unphysical, complex $\eta_H$. For $\eta_V < 3/4$, the inflaton's classical equation of motion has a growing and a decaying solution; we have assumed above motion along the growing attractor solution, as explained in Appendix~\ref{sec:drift_and_diffusion_in_cr}. However, for $\eta_V > 3/4$, the attractor behaviour vanishes and the field starts to oscillate around the minimum. Strictly speaking, our constant roll analysis is not valid in this region, setting the upper limit of validity $\beta = \eta_H < 3/2$. Still, a steep minimum seems more likely to trap the field into eternal inflation than a mild one, so if eternal inflation took place for $\eta_V$ close to $3/4$ ($\beta$ and $\eta_H$ close to $3/2$), we expect eternal inflation also for $\eta_V \geq 3/4$, if other parameter values are kept fixed.\footnote{Naively, the diffusion strength in \eqref{eq:linear_drift_diffusion_drift_cr} seems to diverge in this limit, but this is simply a failure of the approximations used to derive \eqref{eq:linear_drift_diffusion_drift_cr}. See Appendix~\ref{sec:drift_and_diffusion_in_cr} for details and a more accurate result.}

The stochastic constant-roll system is equivalent to that studied earlier in \cite{Tomberg:2023kli}. However, \cite{Tomberg:2023kli} concentrated on stochastic motion on one side of the origin point $\phi=0$ with a hilltop $\eta_V < 0$, since it was later glued to classical evolution that had to end on the right side of the hilltop, and didn't consider boundaries. Here, we have absorbing boundaries on both sides, and we're not particular about the end value of the field, so we may let the field explore both positive and negative $\phi$. We also let $\eta_V$ to be either positive or negative. As discussed in Appendix~\ref{sec:drift_and_diffusion_in_cr}, the stochastic equations of \cite{Tomberg:2023kli} automatically also cover this case. In \cite{Tomberg:2023kli}, the evolution was ended at a fixed $N$, and the dependence on the coarse-graining parameter $\sigma_c$ was cancelled out by varying this final $N$ correspondingly. Since we consider $N\to\infty$, this does not happen, and $\sigma_c$ can play a role: smaller $\sigma_c$ corresponds to a larger patch, experiencing stronger stochastic kicks. However, in the wide limit $\phi_b \gg \sigma$, $\sigma_c$ does not affect conclusions about eternal inflation, which only depends on $\beta = \eta_H$. This limit agrees with the (no-boundary) result of \cite{Tomberg:2023kli}, which found the probability distribution to decay as $e^{-\epsilon_2 N/2} \sim e^{\eta_H N}$ for $\eta_H<0$, matching our discussion below \eqref{eq:lambda_1_asymptotics_beta_negative}.

\subsection{Higher-order drift?}
\label{sec:higher_order_drift}

In the spirit of the previous sections, we might want to study eternal inflation in a model where the drift includes terms up to order $\phi^2$, corresponding to a cubic potential. In fact, such a potential can describe a typical PBH model combining a local minimum with a local maximum, see Figure~\ref{fig:example_potentials}. The eigenvalue equation \eqref{eq:FP_eigenfunctions} becomes one step more complicated. One solution is provided by the triconfluent Heun function, but to the authors' knowledge, the second solution doesn't correspond to a known special function. Both solutions are needed to fix the absorbing boundaries. Making progress requires a full numerical analysis, but numerics quickly run into problems with disparate time scales and stiffness, as we will discuss in the next section. We don't pursue an analysis of $\phi^2$ drift (or higher) in this paper. Fortunately, as discussed in the next section, the linear drift results from Section~\ref{sec:linear_drift} are enough to establish results about eternal inflation in typical PBH models.

\section{Example PBH potentials}
\label{sec:numerical}

\begin{figure}
    \centering
    \includegraphics{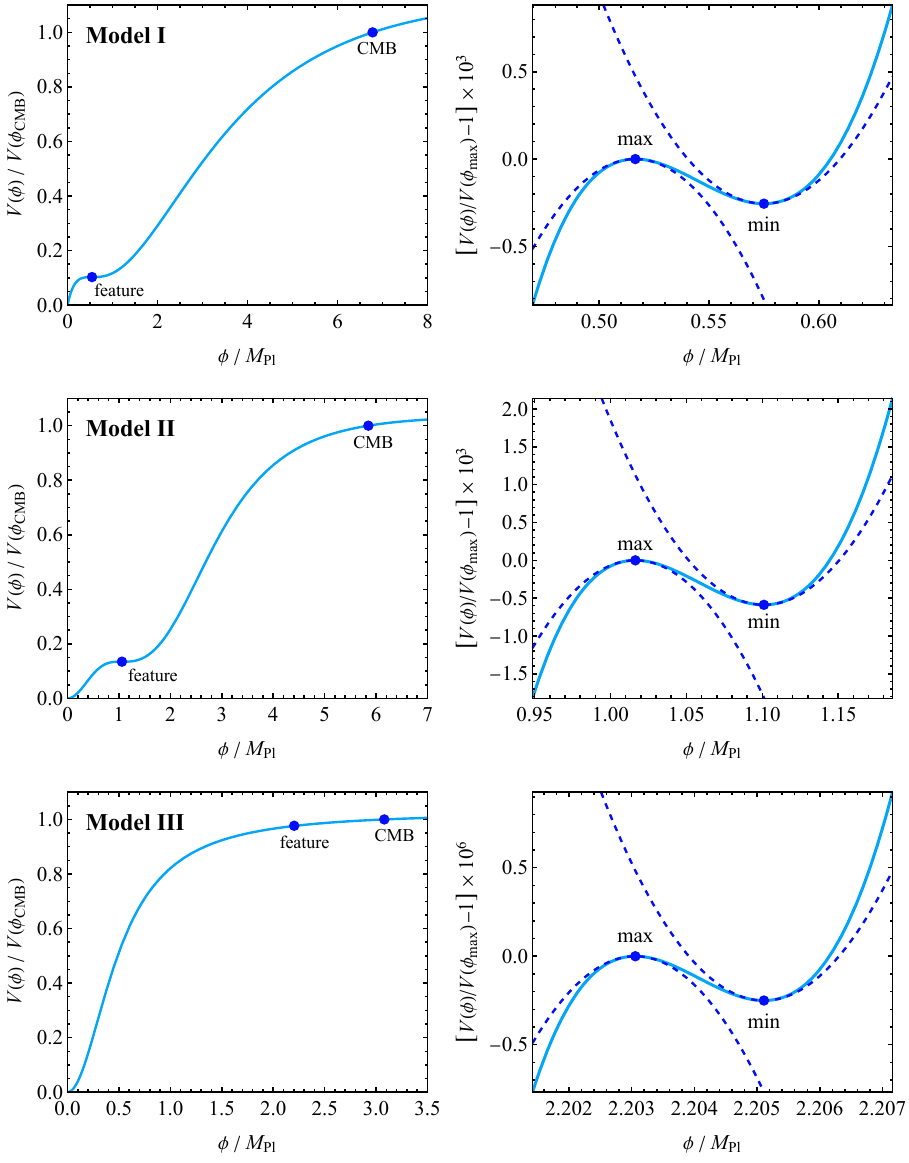}
    \caption{The potentials \eqref{eq:model_I_potential}, \eqref{eq:model_II_potential}, and \eqref{eq:model_III_potential}. The right-hand panels are zoomed into the feature and also depict quadratic approximations around the maximum and minimum (dashed lines).}
    \label{fig:example_potentials}
\end{figure}

Next, we will study eternal inflation in example models. We use the three models analysed in \cite{Karam:2022nym}, drawn from earlier literature; for more similar models, see references within \cite{Karam:2022nym}. All of our potentials feature a flat plateau at large field values, where the CMB perturbations are generated; all models approximately fit the CMB observations. At lower field values, the potentials have a feature consisting of a local minimum followed by a local maximum. The feature produces an enhanced power spectrum that can trigger PBH production in the asteroid mass window. We give the inflationary potentials below and depict them in Figure~\ref{fig:example_potentials}; for more details of their cosmological observables, we refer the reader to \cite{Karam:2022nym} and the original papers \cite{Kannike:2017bxn, Ballesteros:2017fsr, Dalianis:2018frf, Mishra:2019pzq, Ballesteros:2020qam}. We note that in all models, the changes in the potential near the feature are very small compared to the potential's absolute value, justifying the constant-diffusion approximation of Section~\ref{sec:analytical_solutions}.

\paragraph{Model I.} The first potential, `Model I' from \cite{Karam:2022nym}\footnote{There is a typo in \cite{Karam:2022nym}: in their equation (A.2), there shouldn't be factors $1/2$ and $1/3$ in front of the $\sigma^2$ and $\sigma^3$ terms. The potential is reproduced correctly in the current paper. We thank the authors of \cite{Karam:2022nym} for clarifying this.}, is based on a polynomial Jordan frame potential, coupled non-minimally to gravity \cite{Kannike:2017bxn, Ballesteros:2017fsr, Ballesteros:2020qam}. The canonical potential reads
\begin{equation} \label{eq:model_I_potential}
    V(\phi) = \frac{m^2\psi^2(\phi) + \gamma\psi^3(\phi) + \frac{1}{4}\zeta\psi^4(\phi)}{\qty(1+\xi\psi^2(\phi)/\MPl^2)^2} \, ,
\end{equation}
where the function $\psi(\phi)$ must be solved numerically by integrating and inverting
\begin{equation} \label{eq:d_sigma_per_d_phi}
    \frac{\dd \psi}{\dd \phi} = \frac{1+\xi\psi^2/\MPl^2}{\sqrt{1+\frac{\psi^2}{\MPl^2}\qty(\xi + 6\xi^2)}} \, .
\end{equation}
We use the parameter values
\begin{equation} \label{eq:model_I_parameters}
    m^2 = \num{7.7334e-8} \MPl^2 \, , \quad
    \gamma = \num{-1.05442e-6} \MPl \, , \quad
    \zeta = \num{2.10e-5} \, , \quad
    \xi = 80.0118 \, .
\end{equation}
The potential's local maximum, local minimum, and the CMB pivot scale correspond to field values
\begin{equation} \label{eq:model_I_phi_values}
    \phi_\text{max} = 0.516 \, \MPl \, , \quad
    \phi_\text{min} = 0.575 \, \MPl \, , \quad
    \phi_\text{CMB} = 6.78 \, \MPl \, .
\end{equation}

\paragraph{Model II.} The second potential, `Model II' from \cite{Karam:2022nym}, is based on an $\alpha$-attractor, a model with a non-canonical kinetic term \cite{Dalianis:2018frf}. In terms of canonical variables, the potential is
\begin{equation} \label{eq:model_II_potential}
    V(\phi) = V_0\qty[\tanh(\frac{\phi}{\sqrt{6}\MPl}) + A\sin(\frac{1}{f_\sigma}\tanh(\frac{\phi}{\sqrt{6}\MPl}))]^2 \, .
\end{equation}
We use the parameter values
\begin{equation} \label{eq:model_II_parameters}
    V_0 = \num{1.89e-10} \MPl^4 \, , \quad
    A = \num{0.130383} \, , \quad
    f_\sigma = \num{0.129576} \, .
\end{equation}
The potential's local maximum, local minimum, and the CMB pivot scale correspond to field values
\begin{equation} \label{eq:model_II_phi_values}
    \phi_\text{max} = 1.02 \, \MPl \, , \quad
    \phi_\text{min} = 1.10 \, \MPl \, , \quad
    \phi_\text{CMB} = 5.85 \, \MPl \, .
\end{equation}

\paragraph{Model III.} The third potential, `Model III' from \cite{Karam:2022nym}, is a string theory based KKLT construction, with a little `bump' added to it \cite{Mishra:2019pzq}. The potential is
\begin{equation}  \label{eq:model_III_potential}
    V(\phi) = V_0\frac{\phi^2}{\MPl^2/4 + \phi^2}\qty[1 + A\exp(-\frac{(\phi-\phi_0)^2}{2\Delta^2})] \, .
\end{equation}
We use the parameter values
\begin{equation} \label{eq:model_III_parameters}
    V_0 = \num{7.03e-11} \MPl^4 \, , \quad
    A = \num{1.17e-3} \, , \quad
    \phi_0 = 2.188 \MPl \, , \quad
    \Delta = \num{1.59e-2} \, .
\end{equation}
The potential's local maximum, local minimum, and the CMB pivot scale correspond to field values
\begin{equation} \label{eq:model_III_phi_values}
    \phi_\text{max} = 2.2031 \, \MPl \, , \quad
    \phi_\text{min} = 2.2051 \, \MPl \, , \quad
    \phi_\text{CMB} = 3.08 \, \MPl \, .
\end{equation}

\begin{table*}
\begin{center}
\begin{tabular}{L{2.5cm} C{3.2cm} C{3.2cm} C{3.2cm}}
\toprule
& \textbf{Model I} & \textbf{Model II} & \textbf{Model III}  \\
\midrule

$H$ & $\num{4.98e-6} \MPl$ & $\num{3.23e-6} \, \MPl$ & $\num{4.72e-6} \, \MPl$ \\

\midrule

\textbf{Max} & & & \\
$\phi_\text{max}$ & $0.516 \, \MPl$ & $1.02 \, \MPl$ & $2.2031 \, \MPl$ \\
$\phi_b$ & $\num{8.0e-4} \, \MPl$ & $\num{1.2e-3} \MPl$ & $\num{3.0e-5} \, \MPl$ \\
$\eta_V$ & $-0.470$ & $-0.505$ & $-0.365$ \\
$\eta_H$ & $-0.413$ & $-0.441$ & $-0.329$ \\
SR: $\,\phi_b^2/\sigma^2$ & $\num{1.01e6}$ & $\num{5.53e6}$ & $1550$ \\
\phantom{SR:} $\,\lambda_1$ & $\num{0.470}$ & $0.505$ & $0.365$ \\
CR: $\,\phi_b^2/\sigma^2$ & $\num{1.07e4}$ & $\num{4.28e4}$ & $42.3$ \\
\rowcolor{grey} \phantom{CR:} $\,\lambda_1$ & $0.413$ & $0.441$ & $0.329$ \\

\midrule

\textbf{Min} & & & \\
$\phi_\text{min}$ & $0.575 \, \MPl$ & $\num{1.10} \, \MPl$ & $2.2051 \, \MPl$ \\
$\phi_b$ & $\num{0.97e-3} \, \MPl$ & $\num{1.4e-3} \, \MPl$ & $\num{3.2e-5} \, \MPl$ \\
$\eta_V$ & $0.426$ & $0.478$ & $0.349$ \\
$\eta_H$ & $0.514$ & $0.596$ & $0.403$ \\
SR: $\,\phi_b^2/\sigma^2$ & $\num{1.51e6}$ & $\num{6.94e6}$ & $1850$ \\
\phantom{SR:} $\,\lambda_1$ & $\sim0$ & $\sim0$ & $\sim0$ \\
CR: $\,\phi_b^2/\sigma^2$ & $\num{2.70e8}$ & $\num{2.66e9}$ & $\num{1.15e5}$ \\
\rowcolor{grey} \phantom{CR:} $\,\lambda_1$ & $\sim0$ & $\sim0$ & $\sim0$ \\

\midrule

\textbf{Slope} & & & \\
$\phi_b = \frac{\phi_\text{min}-\phi_\text{max}}{2}$ & $0.029 \, \MPl$ & $0.042 \, \MPl$ & $0.0010 \, \MPl$ \\
$\epsilon_V$ & $\num{0.95e-5}$ & $\num{2.41e-5}$ & $\num{7.45e-9}$ \\
SR: $\,\phi_b^2/\sigma^2$ & $\num{1.36e9}$ & $\num{6.78e9}$ & $\num{1.86e6}$ \\
\rowcolor{grey} \phantom{SR:} $\,\lambda_1$ & $\num{1.52e7}$ & $\num{9.11e7}$ & $\num{1.32e4}$ \\

\bottomrule
\end{tabular}
\end{center}
\caption{Model parameters and the lowest eigenvalues $\lambda_1$ in our three example models. The CR eigenvalues (the more accurate ones) are highlighted for the maximum and the minimum; the SR eigenvalue is highlighted for the intermediary slope.}
\label{tab:example_models}
\end{table*}

\paragraph{Eternal inflation.} We expand the potentials around their local maximum and minimum as
\begin{equation} \label{eq:V_expansion_around_extrema}
    V(\phi) = V(\phi_i) + \frac{1}{2}V''(\phi_i)\qty(\phi-\phi_i)^2 + \frac{1}{6}V'''(\phi_i)\qty(\phi-\phi_i)^3 + \dots
\end{equation}
where `$i$' stands for `max' or `min.' Close enough to $\phi_i$, the quadratic term dominates. We set the boundary $\phi_b$ so that when $|\phi-\phi_i| < \phi_b$, then the cubic term is less than one per cent of the quadratic term, giving\footnote{We have also checked that the higher-order terms in \eqref{eq:V_expansion_around_extrema} are subdominant in this regime, each smaller than the previous one.}
\begin{equation} \label{eq:phi_b_from_expansion}
    \phi_b = 0.01\left| \frac{3V''(\phi_i)}{V'''(\phi_i)} \right| \, .
\end{equation}
Then we compute the lowest eigenvalue $\lambda_1$ using the results of Section~\ref{sec:linear_drift}. The relevant model parameters are listed in Table~\ref{tab:example_models}, together with the final $\lambda_1$ values computed with both the slow-roll and the (more accurate) constant-roll approximations.
We note that for all models, the values of $\eta_V$ are moderate, so the SR and CR cases are close to each other. In addition, the $\eta_V$ values are close to each other in the minimum and the maximum, apart from the sign difference.

We also made another estimate of $\lambda_1$ by approximating the potential as a straight line with a fixed slope between the local maximum and the local minimum. This corresponds to the constant drift case of Section~\ref{sec:constant_drift} with
\begin{equation}
    \alpha \approx \frac{|V(\phi_\text{max})- V(\phi_\text{min})|}{|\phi_\text{max} - \phi_\text{min}| V(\phi_\text{max})}
\end{equation}
in the slow-roll approximation. Apart from the extrema, this is the flattest section of the potential (ignoring very large field values $\phi \lesssim \phi_\text{CMB}$) and thus the most prone to eternal inflation.

The results of Table~\ref{tab:example_models} show that all of the models inflate eternally with $\lambda_1 < 3$. Both the extrema and the constant slope are in the wide-range limit of $\phi_b^2/\sigma^2 \gg 1$. As a consequence, the $\lambda_1$ values at the minima are exponentially small, guaranteeing eternal inflation, but the maxima also inflate eternally with $\lambda_1 = |\eta_H| < 3$. Interestingly, the constant slopes have large $\lambda_1$ values with no eternal inflation. Even though the $\epsilon_V$ values on the slopes are small, they are not small enough compared to the diffusion $\sigma^2 \sim H^2$, see the discussion in Section~\ref{sec:constant_drift}. This shows that the parabolic potential analysis of Section~\ref{sec:linear_drift} is crucial for establishing eternal inflation in typical PBH models.

Thus far, we have concentrated on short subsections of our potentials. This turned out to be enough to establish eternal inflation, but it would be interesting to compare these results to numerical solutions of the Fokker--Planck equation over the full field span, say, from the end of inflation $\phi_\text{end}$ to the CMB scale $\phi_\text{CMB}$.\footnote{We still want to restrict the field range from above to exclude extremely large field values where eternal inflation is guaranteed in typical plateau models.} In principle, this is straightforward: we solve the eigenfunction equation \eqref{eq:FP_eigenfunctions} numerically with the initial condition $u(\phi_\text{end}) = 0$, and vary $\lambda_1$ until we find the lowest value that satisfies the other boundary condition $u(\phi_\text{CMB}) = 0$. In the beginning, $u$ tends to grow in a way regulated by the drift term. Since the coefficients in the Fokker--Planck equation change only slowly outside of the feature, this growth can be quasi-exponential. When $\phi$ hits the feature, the drift becomes small, and the conservative $\lambda$ term takes over. It pulls $u$ back towards zero, changing its direction. The drift dictates the later behaviour until the zero-crossing at $\phi_\text{CMB}$.

In practice, finding the numerical solutions is highly non-trivial, since the differential equation is stiff and unstable. This is mainly due to the smallness of $\sigma$, which makes the rate of change of $u$ large. The $\phi$-scales related to $u$'s growth (for most field values) and turn-around behaviour (near the feature) are much smaller than the full $\phi$ span from $\phi_\text{end}$ to $\phi_\text{CMB}$. It is difficult to keep track of $u$ accurately over the whole span, and numerics fail. In fact, we already find numerical instability when trying to solve the system over the field range depicted on the right-hand panels of Figure~\ref{fig:example_potentials} near the feature, not to mention the full potential of the left-hand panels.

As an alternative to solving the eigenfunction equation, one could numerically generate a large number of stochastic field evolutions, employing the system's \emph{Langevin equation} (see, e.g., \cite{Figueroa:2020jkf, Figueroa:2021zah}) and collecting the distribution of first passage times as discussed in Appendix~\ref{sec:first_passage_times}. However, to find the distribution's leading behaviour $e^{-\lambda_1 N}$, the distribution must be resolved so far into the tail that the subleading contributions have decayed away. In Section~\ref{sec:linear_drift} we showed this corresponds to $N \gg 1/(2|\beta|)$ in the wide limit, where $1/(2|\beta|) = 1/(2|\eta_H|) \approx 1$ in the example models. In the full potentials, $N$ here is best understood as the deviation from the mean number of e-folds. For realistic potentials such as those studied here, the probability distribution is extremely suppressed at such large $N$; otherwise, PBHs would be overproduced. Probing this range would require an enormous number of stochastic realisations, making the Langevin approach unfeasible. In a similar semi-analytical computation with explicit time evolution from \cite{Tomberg:2023kli}, the leading behaviour is reached at $N\gtrsim 25$, corresponding to $P\lesssim 10^{-90}$.\footnote{The exponential tail of this model behaves as $e^{-0.4 N}$ and thus undergoes eternal inflation. The same model was considered earlier in \cite{Figueroa:2020jkf, Figueroa:2021zah}, where the behaviour $e^{-32.7 N}$ was reported; this corresponded to smaller $N$ values around $N \sim 1$ when the distribution had not yet settled to the leading large-$N$ behaviour.}

The trouble with numerics shows the usefulness of the sub-potential method and the results of Section~\ref{sec:analytical_solutions}. In fact, we expect the $\lambda_1$-values obtained from the extrema to be close to the full-potential $\lambda_1$, since the extrema dominate the turn-around behaviour of $u$. (In practical terms, of course, the minima already saturate $\lambda_1\sim 0$; the full-potential result can't go below this.)

Based on these results, we claim that \emph{eternal inflation is difficult to avoid in inflection point PBH models.} Indeed, one may ask if it is possible to build an inflection point model that produces significant PBHs but doesn't inflate eternally. To stop eternal inflation near the maximum, one could increase $|\eta_V|$ there. The value of $\eta_V$ at the feature tends to depend on the PBH mass scale, so that higher $|\eta_V|$ correspond to lower PBH masses, see, e.g., \cite{Karam:2022nym}. All three models considered here give PBHs of masses $10^{17}$--$10^{18}$ grams (explaining the similar values of $\eta_V$); PBHs with considerably lower masses and thus higher $|\eta_V|$ values will evaporate by Hawking radiation by today. Eternal inflation near the maximum may not happen for sufficiently low-mass models, but the minimum is still likely to inflate eternally, unless the feature's field range becomes very small so that $\phi_b^2/\sigma^2 \lesssim 1$, see Figure~\ref{fig:linear_drift_eternal_inflation}. The authors don't know if such models exist in the literature.

\section{Consequences of eternal inflation}
\label{sec:discussion}

We have shown that eternal inflation is hard to avoid in typical inflection point models. Let us discuss its consequences.

\subsection{Global structure of the universe}
\label{sec:global_structure}
Typical PBH-forming inflection point models evolve as follows:
\begin{itemize}
    \item At early times, the field follows a slow-roll attractor trajectory. Perturbations are small, matching the CMB observations.
    \item The field encounters the inflection point. Slow roll is broken, overtaken by dual constant-roll phases (the first of which is also called ultra-slow roll) \cite{Karam:2022nym}. The background field has enough energy to pass over the inflection point, but perturbations grow.
    \item The field returns to an attractor with small perturbations. It rolls to the end of inflation, and the universe reheats.
\end{itemize}
Typical perturbations around the background trajectory are small, so the resulting universe is almost homogeneous and isotropic. Large perturbations are rare; they collapse into primordial black holes when they re-enter the Hubble radius \cite{Hawking:1971ei, Carr:1975qj, Green:2020jor, Carr:2025kdk}. The PBH abundance is significant at the inflection point scales, thanks to the enhanced perturbations. The scale determines the PBH mass and formation time. We call the reheated universe outside of the black holes \textbf{U1}.

Eternal inflation arises from extremely large perturbations, which significantly alter the trajectory of the local inflaton field. The new trajectories spend a long time around the inflection point, leading to extra inflationary expansion compared to the typical trajectories. These trajectories are rare, but when weighted by volume, they take over, so that most of the universe inflates at any given time, as discussed in Section~\ref{sec:fokker_planck}. However, when looked from points in the reheated universe U1, these perturbations are simply more extreme versions of primordial black holes; the extra volume is hidden inside, behind an event horizon \cite{Sato:1981bf, Sato:1981gv, Blau:1986cw, Aryal:1987vn, Neeman:1994fxj, Bousso:2006ge}. Due to the large inside volume, the black hole metric's areal radius must be a non-monotonic function of the radial coordinate, corresponding to type II perturbations \cite{Kopp:2010sh, Carr:2014pga, Escriva:2023uko, Harada:2024jxl, Uehara:2024yyp, Escriva:2025eqc, Escriva:2025rja, Inui:2024fgk, Uehara:2025idq}.\footnote{To be more precise, the \emph{perturbations} are type II; the black holes are better categorized by the characteristics of trapping horizons \cite{Uehara:2024yyp, Shimada:2024eec}.} On the inside, space continues to inflate; we call the inside region \textbf{E} (for eternal inflation). Presumably, the barrier between E and U1 looks like a black hole also from E's side, forming a non-traversable wormhole \cite{Kopp:2010sh, Carr:2014pga, Garriga:2015fdk}: observers entering from either side end up in an intermediary singularity. E is a `baby universe' causally disconnected from U1.

While most of the volume of E continues to inflate, individual patches eventually escape the inflection point.\footnote{If eternal inflation happens in a local minimum of the potential, the `classical' field evolution keeps it trapped, and the escaping patches are rare and isolated, in a complete reversal of the behaviour in U1. However, as discussed above, eternal inflation can also happen on a slope that declines towards reheating; classical evolution then drives the field to reheat, and only the volume effect keeps inflation going. The structure of the inflating and reheating domains is then more messy than in the minimum case, but the general picture remains the same. Different phases of eternally inflating spacetimes are discussed in \cite{Sekino:2010vc}.} The escape direction can be either up or down the potential, towards the original CMB regime or towards reheating. Escaping down is favoured by the classical drift and hence more likely, and leads to a reheated universe we call U2; escaping up is more difficult, and leads to eventual reheating into a universe we call U3.

\begin{figure}
    \centering
    \includegraphics{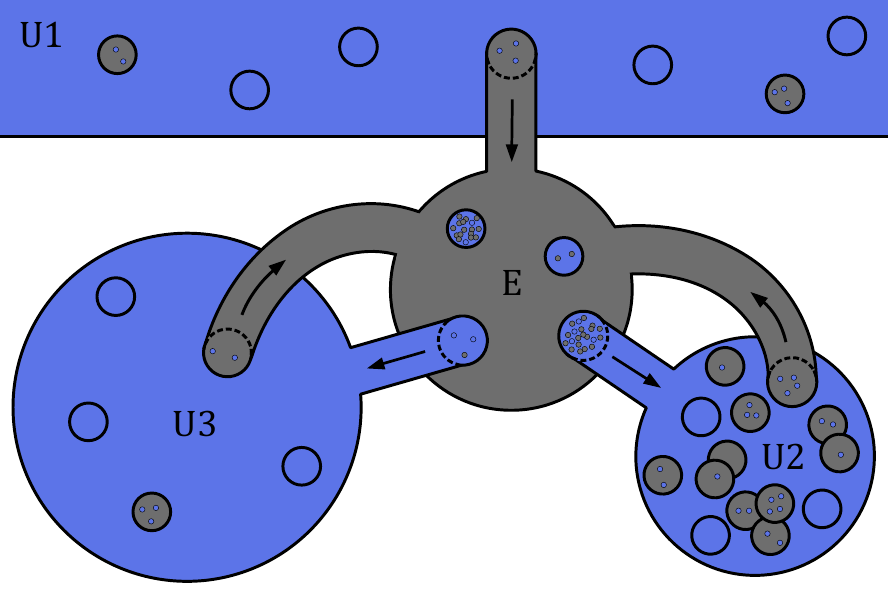}
    \caption{A sketch of the eternal inflation fractal. The grey regions (E) are inflating, the blue regions (U1, U2, U3) have reheated. The circles inside the blue regions are black holes; eternal inflation takes place inside some of them. The arrows indicate the `zoom-in direction' inside the fractal. See main text for further explanation.}
    \label{fig:global_structure}
\end{figure}

The U2 universe is very inhomogeneous on large scales. Just after the field has exited the inflection point, it is easy to diffuse back in, leading to a high abundance of PBHs. At the largest scales, the PBH-forming regions blend together into a complicated non-perturbative structure at the edge of U2 and E.\footnote{This is an extreme version of bubble collisions leaving an imprint on the CMB \cite{Garriga:2006hw, Aguirre:2007an, Aguirre:2009ug, Feeney:2010jj, Kleban:2011pg}.} The scales an observer in U2 might associate with the CMB, 50--60 e-folds before reheating, are not on the original CMB plateau. Instead, these scales are either in the edge region and fully non-perturbative, or arise from a classical trajectory near the hilltop but with a lower field velocity than the original classical trajectory had, and hence higher curvature perturbations.
Since the original trajectory was tuned to produce a peak of $10^{-2}$ in the curvature power spectrum, the spectrum measured in U2 is guaranteed to exceed the observed amplitude of $10^{-9}$.\footnote{In practice, such a classical trajectory is in constant roll near the hilltop \cite{Karam:2022nym}, with a power spectrum scaling as $k^{2\eta_H}$, where the second slow-roll parameter $\eta_H$ from \eqref{eq:eta_H_in_eta_V_CR} is typically of order $-1\ldots -0.1$. The original curvature power spectrum exhibits the same behaviour after its peak; to obtain the U2 power spectrum, one can simply extrapolate this slope to larger scales and higher spectrum values.} U2 does not resemble our observed universe; in fact, it likely collapses quickly due to the abundant PBHs and other inhomogeneities.

In U3, after the initial kick backwards, the field turns around and returns to the original attractor trajectory. Typical patches of space follow the trajectory and reheat into a universe that resembles U1, with U1's curvature power spectrum, except for a transition to non-perturbativity at scales larger than the turnaround point (the edge of the E region). The turnaround point sets a `maximal distance of homogeneity' that varies between different U3 universes. If this distance is pushed outside the observable universe (the field is kicked beyond the original CMB scale), the reheated universe is functionally identical to U1.\footnote{We take here the point of view of an observer like us, 13 billion years after reheating; the non-perturbative scales may eventually enter the observable universe, unless dark energy takes over first.} However, larger distances are less likely, since they require larger perturbations during the exit from E. Most U3 universes are small and, again, incompatible with observations.

As suggested above, the U2 and U3 universes contain inner regions that return to E and look like black holes from the outside. These new E regions spawn new U2 and U3 regions in an endless cycle, as shown in Figure~\ref{fig:global_structure}. An eternally inflating spacetime is a fractal \cite{Aryal:1987vn, Winitzki:2001np, Jain:2019gsq}: zooming in yields more such regions, and a trajectory can pass through any number of phases before eventually hitting reheating. In fact, the separation into E, U2, and U3 is somewhat artificial: the regions blend at the edges in potentially complicated patterns. The separation serves to clarify the discussion and highlight central features of the fractal structure. The blending of the different phases is discussed in \cite{Sekino:2010vc}.

In the literature, two basic types of eternal inflation are considered: the `random-walk type,' where most of the universe rolls down a potential slope but rare patches diffuse up, and `tunneling type,' where most of the universe is stuck in a local minimum but rare patches diffuse out \cite{Linde:1994wt, Bousso:2006ge, Winitzki:2006rn}. The former type is present in typical single-field models at high field values, while the latter is popular in discussions of the string theory landscape \cite{Susskind:2003kw}. Our inflection point models start as random-walk type, with rare patches leaving the classical trajectory, but transitions into the tunneling type if these patches get stuck in a minimum. The models may showcase further eternal inflation (of either type) at scales beyond the CMB region. Extremely large perturbations may even kick the field from the inflection point region back to this `primordial' eternal inflation, which contributes to the same fractal structure as `higher branches.' However, going this far back is rare and doesn't affect our conclusions.

In this paper, we primarily discuss single-field inflection point models of inflation. However, the qualitative discussion here applies to all models with a region capable of eternal inflation between the CMB scales and reheating, including models with multiple fields and more complicated features.

\subsection{Measure problem}
\label{sec:measure_problem}
Above, we demonstrated the existence of two new types of reheated universes, U2 and U3, in addition to the conventional U1. Which of these should we expect to find ourselves in? To answer this question, we need to place a probability measure on different reheated environments. In the spirit of eternal inflation, we want to weight the different regions by their volume -- specifically, volume which can support life and observers like us. However, complications arise in an infinite universe with a fractal structure. This is the \emph{measure problem} \cite{Linde:1994gy, Winitzki:2006rn, Freivogel:2011eg}. The problem is often discussed in the context of string theory vacua when predicting the constants of nature, see, e.g., \cite{Garriga:2005av}; we are interested in predictions for the CMB and large-scale structure (these observables were considered earlier in \cite{Feldstein:2005bm}).

The measure problem has not been fully solved, but most approaches lead to similar conclusions \cite{Winitzki:2006rn}. We have not computed the probabilities in detail for our models\footnote{For quantitative studies of eternally inflating models, see \cite{Aryal:1987vn, Linde:1993xx, Jain:2019gsq, Creminelli:2008es, Sekino:2010vc}.}, but it seems clear\footnote{In our model, all patches end up in the same vacuum, so the measure problem is not as severe as in models with several disconnected vacua \cite{Winitzki:2006rn}. The infinite volume of space still has to be regulated, leading to potential ambiguities, but these shouldn't spoil the picture presented in the text.} that the universe U1 has a small probability compared to U2 and U3, since a large but finite patch of U1 contains an infinite number and volume of inner U2 and U3 patches (replicated by the eternal E region), but U2 and U3 do not contain U1. Furthermore, U2 is more probable than U3, since it is more likely to exit the E region down the potential than up. However, as discussed above, the U2 universes are probably short-lived, unable to produce galaxies and life, so we can exclude them from our considerations by anthropic reasoning.

We are left with U3. We argued above that small U3's are more likely than large ones. We conclude that the most likely place for life like us to occur is in a U3 universe that is as small as possible while still supporting the evolution of an Earth-like planet. It seems clear that large-scale isotropy and homogeneity is not required for this -- a more localized smoothness should be enough. This points to a (probably controversial) conclusion: \emph{eternally inflating PBH models predict highly inhomogeneous large-scale structure}, incompatible with observations.\footnote{This is somewhat reminiscent of the \emph{youngness paradox} \cite{Linde:1994gy, Guth:2007ng}, where young, small universes dominate when probabilities are defined using an equal-time cutoff. However, our conclusion applies even with a proper cutoff \cite{Winitzki:2006rn}.}

Does this rule out such models (including typical inflection point models)? This depends on one's stance on the measure problem. We want to emphasize that the eternally inflating spacetime still contains the U1 region, which produces the conventional predictions for the CMB amplitude and PBH abundance. Since we only observe one universe, one may argue the statistical properties of spacetime as a whole have no bearing on our observations, or that we shouldn't consider ourselves random observers in a simple volume-weighted sense.

At the same time, one of the original motivations for inflation was to explain the homogeneity and isotropy of the observed universe without fine-tuning, from rather generic initial conditions; this motivation is lost, or at least weakened, if homogeneity only applies to rare reheated patches.
In addition, eternal inflation shows that inflation itself is generic, since even extremely rare inflating regions eventually take over the volume \cite{Linde:1986fd}.
If one rejects the statistical relevance of the U2 and U3 universes, one must also reject eternal inflation as a statistical explanation for the onset of inflation.

\section{Discussion}
\label{sec:literature_comparison}

Our results for the eternal inflation conditions (Section~\ref{sec:analytical_solutions}) largely agree with those of \cite{Rudelius:2019cfh}, where the author derived necessary conditions for the potential derivatives for eternal slow-roll inflation. We have extended the analysis to the constant-roll case and carefully considered bounded field intervals.

Around the local maximum, the possibility of eternal inflation was emphasized already in the seminal hilltop inflation paper \cite{Boubekeur:2005zm}. In  \cite{Barenboim:2016mmw}, the authors estimated which $\eta_V$ lead to eternal inflation near a hilltop using a top-hat approximation for the potential. We have improved the analysis by considering a smooth quadratic hilltop, and our $\eta_V$ limits differ from those of \cite{Barenboim:2016mmw} by an order one factor.

Around the local minimum, the decay rate $\Gamma$ -- analogous to our $\lambda_1$ -- can also be computed as quantum tunneling from the Euclidean action. The seminal paper of Hawking and Moss \cite{Hawking:1981fz} found a rate $\Gamma \sim e^{24\pi^2\MPl^4\qty[1/V(\phi_\text{max}) - 1/V(\phi_\text{min})]}$, where $\phi_\text{max}$ is a local maximum and $\phi_\text{min}$ a local minimum of the potential. In \cite{Noorbala:2018zlv}, the authors obtained a compatible result from slow-roll stochastic inflation. Our computation in Section~\ref{sec:linear_drift} is somewhat different, concentrating around the minimum only and using a constant-roll-improved noise. Still, we can use the potential \eqref{eq:linear_drift_V} and $V(\phi_\text{min}) \to V_0$, $V(\phi_\text{max}) \to V(\phi_b)$ to get the estimate $\Gamma \sim e^{-4\pi^2\phi_b^2\eta_V/H^2}$. In the wide limit $\phi_b \gg H$ relevant for typical models, $\Gamma$ is exponentially small signalling eternal inflation, in line with our computation.

When it comes to PBH models, the exponential (and other `heavy') tails of inflationary probability distributions have garnered a lot of attention recently \cite{Pattison:2017mbe, Ezquiaga:2019ftu, Panagopoulos:2019ail, Figueroa:2020jkf, Pattison:2021oen, Biagetti:2021eep, Kitajima:2021fpq, Figueroa:2021zah, Tomberg:2021xxv, Hooshangi:2021ubn, Cai:2021zsp, Achucarro:2021pdh, Tomberg:2022mkt, DeLuca:2022rfz, Ferrante:2022mui, Gow:2022jfb, Pi:2022ysn, Jackson:2022unc, Hooshangi:2023kss, Briaud:2023eae, Tomberg:2023kli, Vennin:2024yzl, Inui:2024sce, Sharma:2024fbr, Miyamoto:2024hin, Animali:2024jiz, Launay:2024qsm, Jackson:2024aoo}. The focus has mostly been on moderate-sized perturbations, which (in an inflection point model) probe the local maximum of the potential; \cite{Atal:2019cdz, Atal:2019erb, Animali:2022otk, Escriva:2023uko} considered cases where the local minimum becomes important.

Eternal inflation was briefly considered in \cite{Animali:2024jiz}, where the authors used volume weighting for perturbation statistics \cite{Tada:2021zzj}. This led to diverging integrals in eternally inflating models. The authors considered quantum-well toy models and excluded eternally inflating cases. Our analysis clarifies the issue: patches with large volumes end up inside black holes, hidden behind an event horizon. An observer in the reheated universe can't see such patches and should exclude them when computing observables. All observers should restrict their attention to their own reheating volume. In practice, this can be achieved with a cutoff in $N$ when computing expectation values.\footnote{Strictly speaking, PBH formation does not depend directly on $N$, but rather on the compaction function \cite{Shibata:1999zs, Harada:2015yda, Musco:2018rwt}, which may be only weakly correlated with $N$ \cite{Raatikainen:2023bzk}. A cutoff in $N$ is a simplification.} See also \cite{Blachier:2025tcq} for a recent discussion on regulating the infinite volumes related to eternal inflation in a plateau potential.

One may ask why eternal inflation has not been considered in detail in PBH models before. One reason may lie in the required e-fold range. Most PBHs form in patches that deviate from the average by a few e-folds at most, while eternal inflation arises from the $N\to\infty$ regime, which is insignificant for PBH statistics (at least in the U1 universe). Computing the $N\to\infty$ tail reliably is also tricky, if the field is allowed to probe any field value; we have sidestepped this issue by concentrating on paths near the extrema of the potential. In addition, eternal inflation is often associated with a curvature power spectrum of order unity, and PBH models keep the spectrum lower, typically at $\sim10^{-2}$. This heuristic reasoning applies in slow-roll inflation, as discussed in Section~\ref{sec:constant_drift}, but fails in inflection point models that break slow roll, in particular, near the local minimum and maximum of the potential. In these models, the possibility of eternal inflation is not apparent from the classical trajectory (though it is obvious from the potential with flat sections); a large deviation from this trajectory is needed to reach the eternally inflating regime. In other words, quantum diffusion doesn't need to dominate on the classical trajectory for eternal inflation to take place -- it is enough for rare perturbations off the classical trajectory to grow strongly in a separate diffusion-dominated regime. Indeed, diffusion domination on the original trajectory typically leads to an overproduction of PBHs \cite{Rigopoulos:2021nhv}, which is not required for eternal inflation.

\section{Conclusions}
\label{sec:conclusions}
We have shown that typical PBH-producing inflection point models inflate eternally, that is, the lowest eigenvalue of their Fokker--Planck equation satisfies $\lambda_1 \leq 3$. We studied three example models from the literature, all of which exhibit wide features, where eternal inflation around the potential's local maximum only requires $\eta_V \geq 6$, and eternal inflation around the local minimum is guaranteed. The eternal behaviour is not obvious from the field's classical average motion, which crosses the inflection point in a finite time, or from typical PBH-forming fluctuations, which represent small deviations from the classical trajectory. However, stronger quantum kicks can push the field off of the classical trajectory, onto a long-lasting attractor near the maximum or into the minimum where it gets stuck. Even though such strong fluctuations are rare, they eventually come to dominate by volume, leading to eternal inflation.

Eternally inflating regions form `baby universes,' type II perturbations separated from surrounding reheated regions by black hole horizons. Inflation goes on inside these extreme primordial black holes. Reheated baby universes dominate the total reheated volume, but they are highly inhomogeneous at large scales, where the marks of the eternally inflating era are visible. If we weight the models' predictions in reheated patches by volume, this makes eternally inflating PBH models incompatible with the CMB and large-scale structure observations.

It is unclear to the authors how serious a threat eternal inflation is for inflection point PBH models. On one hand, CMB-compatible regions still exist in the eternally inflating universe -- they are simply not common if one uses volume weighting. On the other hand, abandoning volume-weighted predictions seems unsatisfactory, since it is needed to argue for the genericity of inflation as an initial condition for cosmic evolution. 

If eternal inflation is a problem, this sets limits on model building. Whether a non-eternal inflection point model can produce a high abundance of PBHs is an open question. In this paper, we have provided tools to check for eternal inflation in such models by giving $\lambda_1$ as a function of the potential parameters in a fixed-width region, where the potential can be approximated as linear or parabolic. These tools can be used in future studies to search for viable non-eternal inflection point models. One should also check for eternal inflation in other PBH-producing models, such as single-field models with different types of features and multi-field setups. This paper's results can serve as a starting point for such checks as well.

It would be interesting to study the fractal structure of the eternally inflating PBH models in more detail. Which fraction of PBHs contains a baby universe, and what is the detailed space-time structure at the interface between the baby universe and the surrounding reheated region? We leave such questions for future study.

\acknowledgments

This work was supported by the Lancaster--Sheffield Consortium for Fundamental Physics under STFC grant: ST/X000621/1.
E.T. is supported by the ``Fonds de la Recherche Scientifique'' (FNRS) under the IISN grant number 4.4517.08.

\appendix

\section{First passage times}
\label{sec:first_passage_times}

In this Appendix, we will define eternal inflation in an alternate way, based on the time the field passes outside of the inflating region.

We start by defining the flux
\begin{equation} \label{eq:j}
    j(\phi, N) \equiv -\partial_\phi\qty(\frac{1}{2} \sigma^2(\phi) P(\phi,N)) - \cV'(\phi) P(\phi,N) \, .
\end{equation}
To interpret $j$, let us consider the time derivative of the survival probability $S(N)$ from \eqref{eq:survival_probability}. Using \eqref{eq:fokker_planck}, we get
\begin{equation} \label{eq:p_to_j}
    \partial_N S(N) =
    \partial_N \int_{\phi_\boundA}^{\phi_\boundB} \dd \phi \, P(\phi, N) = -\int_{\phi_\boundA}^{\phi_\boundB} \dd \phi \, \partial_\phi j(\phi,N) = j(\phi_\boundA, N) - j(\phi_\boundB, N) \, .
\end{equation}
The flux gives the flow of probability through a surface at field value $\phi$ (from small field values to large field values), and \eqref{eq:p_to_j} is the fraction of stochastic paths exiting the bounds $[\phi_\boundA,\phi_\boundB]$ per unit $N$ interval.

In general, $P(\phi,N)$ in \eqref{eq:p_to_j} includes stochastic paths that have passed through the boundaries $\phi_\boundA$, $\phi_\boundB$ multiple times by time $N$. However, if we set the absorbing boundary conditions $P(\phi_\boundA,N)=P(\phi_\boundB,N)=0$, we exclude such paths, including only the first passage of each path into the flux. Each path passes to the boundary sooner or later, and the flux equals the probability density of this \emph{first passage time}, $\PFPT(N)$. We can write it explicitly as
\begin{equation} \label{eq:P_FPT_vs_P}
\begin{aligned}
    \PFPT(N) = -\partial_N S(N)
    &= j(\phi_\boundB, N) - j(\phi_\boundA, N) \\
    &= \frac{1}{2} \qty[\sigma^2(\phi_\boundA) \partial_\phi P(\phi_\boundA,N) - \sigma^2(\phi_\boundB) \partial_\phi P(\phi_\boundB,N)] \, ,
\end{aligned}
\end{equation}
where we emphasize that $P(\phi,N)$ has been solved with the absorbing boundary conditions, which simplifies the expression for $j(\phi_{bi},N)$. With absorbing boundaries at the end of inflation, $\PFPT(N)$ is the probability density for the length of inflation, useful for considerations about eternal inflation.

\subsection{Eternal inflation from first passage times}
\label{sec:eternal_inflation_FPT}

Instead of the uniform-$N$ foliation of Section~\ref{sec:eternal_inflation}, we may consider a foliation where the end-of-inflation (or reheating) points form an equal-time hypersurface. Starting from a finite inflating patch, we say that eternal inflation takes place if the physical volume of this hypersurface diverges \cite{Creminelli:2008es, Tegmark:2004qd}. Volume is again proportional to $e^{3N}$, and the probability density of $N$ at the end of inflation is given by $\PFPT(N)$, so we get
\begin{equation} \label{eq:eternal_inflation_condition_2}
    \text{eternal inflation} \quad \iff \quad \expval{V}_\text{end} \equiv \int_0^\infty e^{3N} \PFPT(N) \dd N = \infty \, .
\end{equation}
Eternal inflation happens if $\PFPT(N)$ does not decrease faster than $e^{-3N}$ as $N \to \infty$.

Since the end-of-inflation surface is well--defined, this description does not suffer from complications related to different time variables, and it is directly linked to the end-of-inflation properties of the Universe, relevant for our existence. We show in Appendix~\ref{sec:sturm-liouville} that our original condition \eqref{eq:eternal_inflation_condition_1} and \eqref{eq:eternal_inflation_condition_2} lead to identical conclusions about eternal inflation. For this, we need to first study the time evolution of $\PFPT(N)$.

\subsection{Adjoint Fokker--Planck equation}
\label{sec:adjoint}
In the main text, we didn't pay close attention to the initial conditions for the field distribution $P(\phi,N)$. Let us remedy this and consider stochastic evolutions that start from fixed $\phi=\phi_0$ at time $N=N_0$, and denote the probability distribution of $\phi$ at a later time $N$ by $\Pcond{\phi}{N}{\phi_0}{N_0}$. This quantity follows the Fokker--Planck equation w.r.t. the primary arguments $\phi, N$ -- but what about the secondary arguments $\phi_0, N_0$?

By the completeness of probabilities, we must have \cite{Vennin:2024yzl}
\begin{equation} \label{eq:p_completeness}
    \Pcond{\phi_2}{N_2}{\phi_0}{N_0} =
    \int \dd \phi_1 \Pcond{\phi_2}{N_2}{\phi_1}{N_1} \Pcond{\phi_1}{N_1}{\phi_0}{N_0}
\end{equation}
where $N_0 < N_1 < N_2$ -- that is, each path from $\phi_0$ to $\phi_2$ must pass through some field value $\phi_1$ at an intermediary time $N_1$. Taking a derivative with respect to $N_1$, using the Fokker--Planck equation \eqref{eq:fokker_planck}, and integrating by parts, we get\footnote{\label{fn:boundaries} In partial integration, we assume that the boundary terms vanish. If the boundary is at infinity, this applies for all reasonable probability distributions. Otherwise, this determines the boundary conditions of $\Pcond{\phi}{N}{\phi_0}{N_0}$ with respect to $\phi_0$ in terms of boundary conditions with respect to $\phi$. For absorbing boundary conditions $\Pcond{\phi_b}{N}{\phi_0}{N_0} = 0$, the relationship is easy: the boundary terms vanish if (and only if) also $\Pcond{\phi}{N}{\phi_b}{N_0} = 0$, for all $N$ and $N_0$ (in other words, the field can't escape from the absorbing boundary).}
\begin{equation} \label{eq:to_adjoint}
\begin{aligned}
    0 &= \int \dd \phi_1 \partial_{N_1}\Pcond{\phi_2}{N_2}{\phi_1}{N_1} \times \Pcond{\phi_1}{N_1}{\phi_0}{N_0} \\
    &\qquad + \Pcond{\phi_2}{N_2}{\phi_1}{N_1} \times \partial_{N_1}\Pcond{\phi_1}{N_1}{\phi_0}{N_0} \\
    &= \int \dd \phi_1 \partial_{N_1}\Pcond{\phi_2}{N_2}{\phi_1}{N_1} \times \Pcond{\phi_1}{N_1}{\phi_0}{N_0} \\
    &\qquad+ \Pcond{\phi_2}{N_2}{\phi_1}{N_1} \times \partial_{\phi_1}\qty[\partial_{\phi_1}\qty(\frac{1}{2}\sigma^2(\phi_1)\Pcond{\phi_1}{N_1}{\phi_0}{N_0}) + \cV'(\phi_1) \Pcond{\phi_1}{N_1}{\phi_0}{N_0}] \\
    &= \int \dd \phi_1 \qty[\partial_{N_1} \Pcond{\phi_2}{N_2}{\phi_1}{N_1} + \frac{1}{2}\sigma^2(\phi_1)\partial^2_{\phi_1}\Pcond{\phi_2}{N_2}{\phi_1}{N_1} - \cV'(\phi_1) \partial_{\phi_1}\Pcond{\phi_2}{N_2}{\phi_1}{N_1}] \\
    &\qquad \times \Pcond{\phi_1}{N_1}{\phi_0}{N_0} \, ,
\end{aligned}
\end{equation}
so we conclude
\begin{equation} \label{eq:adjoint_fokker_planck_pre}
    \partial_{N_1} \Pcond{\phi_2}{N_2}{\phi_1}{N_1} = -\frac{1}{2}\sigma^2(\phi_1)\partial^2_{\phi_1}\Pcond{\phi_2}{N_2}{\phi_1}{N_1} + \cV'(\phi_1) \partial_{\phi_1}\Pcond{\phi_2}{N_2}{\phi_1}{N_1} \, .
\end{equation}
Let us recall the Fokker--Planck operator from \eqref{eq:fokker_planck} and define its \emph{adjoint}\footnote{When $\cL$ and operates on a function, multiplication with the right hand factors in the terms takes place before differentiation.}
\begin{equation} \label{eq:L_FP}
    \cL_{\text{FP},\phi} = \frac{1}{2}\partial^2_{\phi}\sigma^2(\phi) + \partial_\phi \cV'(\phi) \, ,  \qquad
    \cLd_{\text{FP},\phi} = \frac{1}{2}\sigma^2(\phi)\partial^2_{\phi} - \cV'(\phi) \partial_\phi \, ,
\end{equation}
where the adjoint is taken with respect to the inner product $f\cdot g = \int \dd \phi f(\phi) g(\phi)$ -- that is, $f \cdot \cL g = \int \dd \phi f(\phi) \cL g(\phi) = \int \dd \phi \qty[\cLd f(\phi)] g(\phi) = \cLd f \cdot g$. Then
\begin{equation} \label{eq:fokker_planck_in_L}
    \partial_N \Pcond{\phi}{N}{\phi_0}{N_0} = \cL_{\text{FP},\phi} \Pcond{\phi}{N}{\phi_0}{N_0} \, , \qquad
    \partial_{N_0} \Pcond{\phi}{N}{\phi_0}{N_0} = -\cLd_{\text{FP},\phi_0} \Pcond{\phi}{N}{\phi_0}{N_0} \, .
\end{equation}
Finally, since there is no explicit time dependence in the system, $\Pcond{\phi}{N}{\phi_0}{N_0}$ can only depend on the difference $N-N_0$. As a consequence,
\begin{equation} \label{eq:N_vs_N0_derivatives}
    \partial_N \Pcond{\phi}{N}{\phi_0}{N_0} = -\partial_{N_0} \Pcond{\phi}{N}{\phi_0}{N_0} \implies \cL_{\text{FP},\phi} \Pcond{\phi}{N}{\phi_0}{N_0} = \cLd_{\text{FP},\phi_0} \Pcond{\phi}{N}{\phi_0}{N_0} \, .
\end{equation}

Equipped with these results, we can say something more about the first passage time distribution $\PFPT$. Let us expand $\PFPT$ to a function of not only $N$ but also of $\phi$, which now denotes the initial field value from which the stochastic evolution starts. With equation \eqref{eq:P_FPT_vs_P}, we obtain\footnote{We omit the integration bounds for generality; they don't affect the argument.}
\begin{equation} \label{eq:P_FPT_evolution}
\begin{aligned}
    \partial_N \PFPT(N,\phi)
    &= -\partial^2_N \int \Pcond{\tilde{\phi}}{N}{\phi}{N_0} \dd \tilde{\phi}
    = -\partial_N \int \cL_{\text{FP},\tilde{\phi}} \Pcond{\tilde{\phi}}{N}{\phi}{N_0} \dd \tilde{\phi} \\
    &= -\partial_N \int \cLd_{\text{FP},\phi} \Pcond{\tilde{\phi}}{N}{\phi}{N_0} \dd \tilde{\phi}
    = \cLd_{\text{FP},\phi} \PFPT(N,\phi) \, .
\end{aligned}
\end{equation}
This is the \emph{adjoint Fokker--Planck equation} \cite{Vennin:2015hra, Pattison:2017mbe, Ezquiaga:2019ftu}, a partial differential equation for $\PFPT(N,\phi)$ similar to \eqref{eq:fokker_planck}.\footnote{\label{fn:FPT_intial_conditions}Instead of a single initial field value, we may wish to consider a distribution $P(\phi,N_0)$ in initial $\phi$ for a first passage time problem. The full $\PFPT(N)$ can then be obtained as the weighted integral $\int \dd \phi \PFPT(N,\phi)P(\phi,N_0)$. Importantly, this does not change the considerations related to the eigenvalues below.} Let us write it down explicitly:
\begin{equation} \label{eq:adjoint_fokker_planck}
        \partial_{N} \PFPT(N,\phi) = \frac{1}{2}\sigma^2(\phi_1)\partial^2_{\phi}\PFPT(N,\phi) - \cV'(\phi) \partial_{\phi}\PFPT(N,\phi) \, .
\end{equation}
The late-time behaviour of $\PFPT$ can then be analysed similarly to that of the original distribution $P$, through an eigenfunction expansion, as we discuss in detail in Appendix~\ref{sec:sturm-liouville}.

\section{Sturm--Liouville theory}
\label{sec:sturm-liouville}
In a Sturm--Liouville problem (see, e.g., \cite{arfken2011mathematical, sturmliouville}), one seeks values $\lambda$ and functions $y(x)$ that solve the differential equation
\begin{equation} \label{eq:sturm_liouville_equation}
    \frac{\dd}{\dd x}\qty[p(x)\frac{\dd}{\dd x}y(x)] + q(x)y(x) = -\lambda w(x)y(x) \, , \quad x \in [x_1,x_2] \, .
\end{equation}
The problem is called \emph{regular} if $p(x)$, $q(x)$, and $w(x)$ are sufficiently well-behaved real functions, $p(x)>0$ and $w(x)>0$, the interval $[x_1,x_2]$ is finite, and the boundary conditions 
\begin{equation} \label{eq:sturm_liouville_boundaries}
    \alpha_1 y(x_1) + \beta_1 y'(x_1) = 0 \, , \qquad
    \alpha_2 y(x_2) + \beta_2 y'(x_2) = 0
\end{equation}
apply for some $\alpha_1$, $\alpha_2$, $\beta_1$, and $\beta_2$, where at least one of $\alpha_i$ and $\beta_i$ is non-zero. A regular Sturm--Liouville problem has a discrete, numerable set of real solutions $y_n(x)$, with corresponding real $\lambda_n$, so that $n \in \mathbb{Z}_+=1,2,3,\dots$ and
\begin{itemize}
    \item $\lambda_1 < \lambda_2 < \dots \to \infty$,
    \item $y_n(x)$ has exactly $n-1$ zeroes in the interval $]x_1,x_2[$,
    \item the functions $y_n(x)$ form a complete basis for all (sufficiently regular) functions of $x \in [x_1,x_2]$ that obey the boundary conditions \eqref{eq:sturm_liouville_boundaries}, and the basis is orthonormal with respect to the inner product $f \overset{w}{\circ} g = \int_{x_1}^{x_2}\dd x f(x)g(x)w(x)$.
\end{itemize}

To compare the Sturm--Liouville problem to our Fokker--Planck equation, let us define the Sturm--Liouville operator
\begin{equation} \label{eq:L_SL}
    \cL_{\text{SL},x} = \frac{\dd}{\dd x}p(x)\frac{\dd}{\dd x} + q(x) \quad\implies\quad \cL_{\text{SL},x}y(x) = -\lambda w(x)y(x) \, .
\end{equation}
Note that $\cL_{\text{SL},x}$ is self-adjoint, that is, $\cL_{\text{SL},x} = \cLd_{\text{SL},x}$ in the sense discussed below \eqref{eq:L_FP}. Let us switch the variable from $x$ to $\phi$ and choose
\begin{equation} \label{eq:choosing_p_q_w}
\begin{gathered}
    w(\phi) = \sigma^2(\phi)\exp(\int_{\phi_\boundA}^\phi \frac{2\cV'(\tilde{\phi})}{\sigma^2(\tilde{\phi})}\dd \tilde{\phi}) \, , \\
    p(\phi) = \frac{1}{2}\sigma^2(\phi)w(\phi) \, , \quad
    q(\phi) = \frac{1}{2}w(\phi)\partial^2_\phi\qty[\sigma^2(\phi)] + w(\phi)\cV''(\phi) \, .
\end{gathered}
\end{equation}
With these choices, we have
\begin{equation} \label{eq:FP_vs_SL_operator}
    \cL_{\text{SL},\phi} = w(\phi)\cL_{\text{FP},\phi} \, ,
\end{equation}
and the two operators share eigenvalues and functions:
\begin{equation} \label{eq:FP_vs_SL_operator_eigenstates}
    \cL_{\text{SL},\phi} u(\phi) = -\lambda w(\phi)u(\phi) \quad\iff\quad
    \cL_{\text{FP},\phi} u(\phi) = -\lambda u(\phi) \, .
\end{equation}
Furthermore, since $\cL_{\text{SL},\phi}$ is self-adjoint, we have\footnote{As an operator, $w(\phi)$ simply multiplies the target function and is thus self-adjoint; its inverse is just multiplication with $1/w(\phi)$.} 
\begin{equation} \label{eq:FP_vs_adjoint_via_SL}
    w(\phi)\cL_{\text{FP},\phi} = \qty[w(\phi)\cL_{\text{FP},\phi}]^\dagger = \cLd_{\text{FP},\phi}w(\phi) \implies \cLd_{\text{FP},\phi} = w(\phi)\cL_{\text{FP},\phi}\qty[w(\phi)]^{-1} \, .
\end{equation}
This lets us write\footnote{From here, one can also figure out the relationship between the boundary conditions of the functions operated on by $\cL_{\text{FP},\phi}$ and $\cLd_{\text{FP},\phi}$. For absorbing boundary conditions, all the functions must vanish at the boundary, see also footnote \ref{fn:boundaries}.}
\begin{equation} \label{eq:FP_vs_adjoint_eigenstates}
    \cL_{\text{FP},\phi} u(\phi) = -\lambda u(\phi) \quad\iff\quad
    \cLd_{\text{FP},\phi} \bar{u}(\phi) = -\lambda \bar{u}(\phi) \, , \quad \bar{u}(\phi) = w(\phi)u(\phi) \, .
\end{equation}
In other words, $\cL_{\text{FP},\phi}$ and $\cLd_{\text{FP},\phi}$ share eigenvalues, and their eigenfunctions are related by a factor of $w(\phi)$. Furthermore, the orthogonality statement of the Sutrm--Liouville solutions becomes an orthogonality statement for the two sets of eigenfunctions:
\begin{equation} \label{eq:orthogonality_SL_vs_FP}
    u_1 \overset{w}{\circ} u_2 = 0 \iff
    0 = \int \dd\phi \, u_1(\phi)u_2(\phi) w(\phi)
    = \int \dd\phi \, \bar{u}_1(\phi)u_2(\phi) = \bar{u}_1 \cdot u_2 \, .
\end{equation}
The fact that adjoint operators share eigenvalues with eigenvectors that are orthogonal is not surprising -- it follows from the general theory of Hilbert spaces \cite{sturmliouville} -- but Sturm--Liouville theory helps us say more about the eigenvalues, if the regularity conditions are satisfied. In particular, the theory guarantees the existence of a lowest eigenvalue $\lambda_1$. The theory also identifies the corresponding eigenfunction with the unique function that satisfies the Sturm--Liouville equation and the boundary conditions and has no zeroes inside the domain. In principle, this can be searched for numerically using a shooting method.

The eigenvalue decomposition of equations \eqref{eq:P_expanded_in_u} and \eqref{eq:P_expanded_in_u_evolution} also applies for $\PFPT(N,\phi)$, when we replace $\cL_{\text{FP},\phi}$ by $\cLd_{\text{FP},\phi}$ and $u_n(\phi)$ by $\bar{u}_n(\phi)$. The eternal inflation conditions \eqref{eq:eternal_inflation_condition_1}, \eqref{eq:eternal_inflation_condition_2} are then equal, as claimed above, both yielding equation~\eqref{eq:eternal_inflation_condition_in_lambda}, $\lambda_1 \leq 3$.
Let us emphasize that $\lambda_1$ may be solved from either the Fokker--Planck or the adjoint Fokker--Planck equation, since these share eigenvalues.

Let us comment on the possible values of $\lambda_1$.
In general, we don't expect the integral over $P(\phi, N)$ to increase in time -- this would correspond to probability flowing in from the outside, and is forbidden by any reasonable boundary conditions. This sets the restriction $\lambda_1 \geq 0$.\footnote{This limit can also be derived explicitly by making a slightly different transformation to both $P$ and $\phi$, see \cite{Helmer:2006tz, Winitzki:1995pg}.} If the boundary conditions are set to infinity and the potential is bounded from below, or if we have reflecting boundaries (flux $j$ zero at the boundary), then there is a stationary configuration, corresponding to the limiting case $\lambda_1 = 0$. It is easy to see that the corresponding eigenfunction of $\cLd_{\text{FP},\phi}$ from \eqref{eq:L_FP} is simply a constant (note that $\cLd_{\text{FP},\phi}$ doesn't represent a first passage process with these boundary conditions); the corresponding eigenfunction of $\cL_{\text{FP},\phi}$ is then proportional to $\qty[w(\phi)]^{-1}$. This gives us a physical interpretation for $\qty[w(\phi)]^{-1}$: it is the time-independent equilibrium distribution (indeed, one can find the functional form \eqref{eq:choosing_p_q_w} also from the requirement $j(\phi,N)=0$ with $P(\phi,N)=\qty[w(\phi)]^{-1}$). In all cases with absorbing boundaries, some probability is always flowing out of bounds, implying $\lambda_1 > 0$.

The lowest eigenfunction $u_1(\phi)$ can also be interpreted as a stationary solution for the eternally inflating universe \cite{Nambu:1988je, Nambu:1989uf, Linde:1993xx, Linde:1993nz, Garcia-Bellido:1994gng}: it gives the asymptotic field distribution in the eternal regime. The lowest eigenvalue $\lambda_1$ is related to the fractal dimension of the eternally inflating universe \cite{Aryal:1987vn, Winitzki:2001np}: $d=3-\lambda_1$.

\section{Eternal inflation from a sub-potential}
\label{sec:eternal_inflation_from_subpotential}
In Section~\ref{sec:sub_potential}, we showed that eternal inflation in a subsection of the potential leads to eternal inflation in the full potential, if we start within the sub-interval and if the lowest eigenvalues exist for both intervals. The last assumption is true if both intervals are finite, so that the Sturm--Liouville problem is regular. In this Appendix, we prove the same result more generally, with arbitrary initial conditions and for a semi-infinite field range, without referring to the eigenvalues.

\paragraph{Volume at a fixed time.}
Let us first consider the expectation value of an inflating volume at a fixed time $N$ (see \eqref{eq:eternal_inflation_condition_1} and \eqref{eq:survival_probability}):
\begin{equation} \label{eq:V_N}
    \expval{V}_N = e^{3N} \int_{\phi_\boundA}^{\phi_\boundB} P(\phi,N) \dd \phi = e^{3N}S(N) \, .
\end{equation}
Like above, we assume inflation takes place within the interval $]\phi_\boundA,\phi_\boundB[$ and solve $P(\phi,N)$ from the Fokker--Planck equation with the desired initial condition at $N=0$ and absorbing boundary conditions at the edges. One of the edges may be infinitely far away, but at least one is finite, so that probability keeps leaking out, and $S(N_1) > S(N_2)$ for $N_1 < N_2$. We may say that eternal inflation takes place if the leakage is not fast enough to overcome the $e^{3N}$ growth factor.

We can discuss the leaking probability in terms of individual stochastic paths. Each path carries equal probability weight; when a path hits a boundary, it is absorbed and removed from the field interval $]\phi_\boundA,\phi_\boundB[$. The integral \eqref{eq:V_N} sums over the remaining paths, giving the fraction of paths still within the interval. We can find a lower bound for the integral by considering only a subset of the surviving paths. In particular, let us consider the subset that is within a more restricted interval $]\phi_\subboundA,\phi_\subboundB[$ at some finite time $\Delta N_a$ and stays within the interval until time $N$.
For $N>\Delta N_a$, these paths have a probability distribution $\tilde{P}(N,\phi)$, solved from Fokker--Planck with absorbing boundaries at $\phi_\subboundA$ and $\phi_\subboundB$ to remove paths that would venture outside the interval, and with the initial condition $\tilde{P}(\phi,\Delta N_a) = P(\phi,\Delta N_a)$ for $\phi \in 
 \, ]\phi_\subboundA,\phi_\subboundB[$. Since these are a subset of all paths, we get
\begin{equation} \label{eq:P_vs_tilde_P}
\begin{gathered}
    P(\phi, N) > \tilde{P}(\phi, N) \, , \quad \phi \in \, ]\phi_\subboundA,\phi_\subboundB[ \, , \quad N>\Delta N_a \\
    \implies
    \underbrace{e^{3N}\int_{\phi_\boundA}^{\phi_\boundB} P(\phi,N)\dd\phi}_{=\expval{V}_N} >
    \underbrace{e^{3N}\int_{\phi_\subboundA}^{\phi_\subboundB} \tilde{P}(\phi,N)\dd\phi}_{\equiv \exptV_N} \quad \text{for large $N$} \, ,
\end{gathered}
\end{equation}
where we also used $]\phi_\subboundA,\phi_\subboundB[ \, \subset \, ]\phi_\boundA,\phi_\boundB[$ and the positivity of probability densities to restrict the domain of integration. If we define eternal inflation as the divergence (or at least non-vanishing) of $\expval{V}_N$ at late times, we see that eternal inflation in the sub-potential implies eternal inflation in the full potential: $\exptV_N \to \infty \implies \expval{V}_N \to \infty$ as $N\to \infty$ (similarly,  $\exptV_N >0 \implies \expval{V}_N >0$).

\paragraph{Volume at the end of inflation.}
Let us then consider the expectation value of the volume at the end-of-inflation hypersurface, where the paths get absorbed. Using the first passage time distribution of Appendix~\ref{sec:first_passage_times}, we have (see \eqref{eq:eternal_inflation_condition_2})
\begin{equation}
    \expval{V}_\text{end} = \int_0^\infty e^{3N}\PFPT(N) \dd N \, ,
\end{equation}
where $\PFPT(N)$ is solved from the adjoint Fokker--Planck equation with the appropriate initial conditions and absorbing boundaries at $\phi_\boundA$ and $\phi_\boundB$. The value $\PFPT(N)\dd N$ gives the fraction of paths for which inflation ends in a bin of width $\dd N$ around $N$. We can again find a lower bound for this by only considering a subset of such paths.

This time, our subset consists of paths that are within the interval $]\phi_\subboundA,\phi_\subboundB[$ at time $\Delta N_a$ (as above), stay there until they hit either $\phi_\subboundA$ or $\phi_\subboundB$ in an interval $\dd N$ around time $N - \Delta N_b$, and then drift to one of the outer boundaries $\phi_\boundA$ or $\phi_\boundB$, hitting it within interval $\dd N$ of time $N$.
Let $\tPFPT(N)$ denote the FPT distribution from $]\phi_\subboundA,\phi_\subboundB[$ to either $\phi_\subboundA$ or $\phi_\subboundB$, assuming an initial distribution at time $\Delta N_a$ set by the full stochastic process (see footnote~\ref{fn:FPT_intial_conditions} on how to solve this from the adjoint Fokker--Planck equation). We then get
\begin{equation} \label{eq:PFPT_vs_PFTP_tilde_pre}
\begin{aligned}
    \PFPT(N) &> \tilde{P}_{\text{FPT},\tilde{b}1}(N - \Delta N_b) \dd N \times \PFPT\qty(\Delta N_b, \phi_\subboundA) \\
    &+ 
    \tilde{P}_{\text{FPT},\tilde{b}2}(N - \Delta N_b) \dd N \times \PFPT\qty(\Delta N_b, \phi_\subboundB) \, ,
\end{aligned}
\end{equation}
where $\tilde{P}_{\text{FPT},\tilde{b}i}$ is the FPT probability through the inner boundary $\phi_{\tilde{b}i}$, $i=1,2$, so that the full $\tPFPT = \tilde{P}_{\text{FPT},\tilde{b}1} + \tilde{P}_{\text{FPT},\tilde{b}2}$; $\PFPT\qty(\Delta N_b, \phi_{\tilde{b}i})$ is the FPT probability in the full potential after $\Delta N_b$ e-folds, starting from $\phi_{\tilde{b}i}$, as per our usual notation. Writing $m(\Delta N_b) \equiv \min[\PFPT\qty(\Delta N_b, \phi_\subboundA), \PFPT\qty(\Delta N_b, \phi_\subboundB)]$, we further get
\begin{equation} \label{eq:PFPT_vs_PFTP_tilde}
\begin{gathered}
    \PFPT(N) > m(\Delta N_b) \tPFPT(N - \Delta N_b) \, , \quad N>\Delta N_a \\
    \implies
    \underbrace{\int_0^\infty e^{3N}\PFPT(N)\dd N}_{=\expval{V}_\text{N}} >
    m(N_b)\underbrace{\int_{\Delta N_a}^\infty e^{3N}\tPFPT(N - \Delta N_b)\dd N}_{\equiv \exptV_\text{end}} \, ,
\end{gathered}
\end{equation}
and we again used the positivity of the probabilities to limit the integration range. We note that the finite quantities $\Delta N_a$ and $\Delta N_b$ make no difference for the finiteness of the last integral.
If we define eternal inflation as the divergence of $\expval{V}_\text{end}$, then, once again, eternal inflation in the sub-potential implies eternal inflation in the full potential: $\exptV_\text{end} \to \infty \implies \expval{V}_\text{end} \to \infty$ as $N\to \infty$.

Note that both our $\tilde{P}$ and $\tPFPT$ had non-trivial initial conditions, arising from the stochastic evolution until the time $\Delta N_a$. If the functions can be decomposed into sums of the form \eqref{eq:P_expanded_in_u} with \eqref{eq:P_expanded_in_u_evolution}, the initial conditions don't matter: at late times, the leading exponent $\tilde{\lambda}_1$ dominates, and $\tilde{\lambda}_1 \leq 3$ is a sufficient condition for eternal inflation also in the full potential.

\section{\boldmath Properties of the $g$ function}
\label{sec:g_properties}
Let us study the properties of the function $g(x)$, defined in \eqref{eq:g_defined}. We first generalize the function to $g_n(x)$, where $n \in \mathbb{Z}_+={1,2,3,\dots}$ enumerates the various solutions of \eqref{eq:g_defined} in order,
\begin{equation} \label{eq:g_n_defined}
    \iFi\qty(g_n(x); \frac{1}{2}; -x) = 0 \, , \quad k < l \iff g_k(x) < g_l(x) \, .
\end{equation}
The $g(x)$ of the main text is the lowest solution, $g(x)\equiv g_1(x)$. The higher $g_n(x)$ give the higher eigenvalues $\lambda_n$ with the replacement $g(x) \to g_n(x)$ in \eqref{eq:lambda_in_g_beta_negative}, \eqref{eq:lambda_in_g_beta_positive} (assuming symmetric initial conditions, so that only the even branch of \eqref{eq:kummer_solved} contributes).

Let us first consider small (positive) values of $x$. We first write \eqref{eq:g_n_defined} in terms of the series representation \eqref{eq:hypergeometric_series}:
\begin{equation} \label{eq:g_eq_series}
    \sum_{k=0}^{\infty} \frac{g_n(x)^{(k)} (-x)^k}{(1/2)^{(k)}k!} = 0 \, .
\end{equation}
In the series, we only need to keep the leading order contributions in the small $x$. However, even though $x$ is small, $g(x)$ may be large, and its largeness may cancel out the smallness of $x$ in some terms. To ensure we keep all relevant terms, we'll re-arrange the series into powers of $g_n(x)$, and for each power of $g(x)$, we'll only keep the leading $x$ contribution. The power $g_n(x)^k$ appears in \eqref{eq:g_eq_series} in all terms starting from the $k$th; in this $k$th term, it is accompanied by the smallest power of $x$, that is, $x^k$. Only keeping these terms, \eqref{eq:g_eq_series} simplifies to
\begin{equation} \label{eq:g_eq_series_simplified}
\begin{gathered}
    \sum_{k=0}^{\infty} \frac{\qty[-g_n(x)x]^k}{(1/2)^{(k)}k!}
    \equiv \oFi\qty(;\frac{1}{2};-g_n(x)x)
    = \cos(\sqrt{4g_n(x)x}) = 0 \\
    \implies
    g_n(x) = \frac{\pi^2}{4x}\qty(n-\frac{1}{2})^2
    \, , \quad n \in \mathbb{Z}_+ \, , \quad
    0 < x \ll 1 \, .
\end{gathered}
\end{equation}
Here $\oFi$ is defined analogously to  $\iFi$, but with no rising factorial in the numerator. Its simplification into a cosine can be verified term by term in the series expansion. Choosing the lowest solution, we get
\begin{equation} \label{eq:g_for_small_x}
    g(x) = \frac{\pi^2}{16x} \, , \qquad 0<x\ll 1 \, , 
\end{equation}
as we claimed above in \eqref{eq:g_asymptotics}. We note that $g(x)$ indeed diverges for small $x$, as we anticipated earlier.

What about large $x$? Let us study the alternate form of \eqref{eq:g_eq_series} given by Kummer's transformation \eqref{eq:kummers_transformation}:
\begin{equation} \label{eq:g_n_alternative}
    e^{-x}\sum_{k=0}^{\infty} \frac{\qty[\frac{1}{2} - g_n(x)]^{(k)} x^k}{(1/2)^{(k)}k!} = 0 \, .
\end{equation}
The exponential prefactor brings the left-hand side to zero for $x \to \infty$ if the sum doesn't grow too fast. This is guaranteed if
\begin{equation} \label{eq:g_n_for_large_x}
    g_n(x) \xrightarrow{x\to\infty} n - \frac{1}{2} \, , \quad n \in \mathbb{Z}_+ \, ,
\end{equation}
since then the rising factorials in the numerator contain a factor $0$ starting from $k=n$, killing all higher-order terms and turning the sum into a polynomial of order $n-1$. A numerical check confirms this is indeed the full set of asymptotic solutions. In particular, the lowest solution $g(x)=g_1(x) \to 1/2$, matching \eqref{eq:g_asymptotics}, and the left-hand side of \eqref{eq:g_n_alternative} reduces simply to $e^{-x}$ (this is also clear from \eqref{eq:g_eq_series}, where the rising factorials cancel).

To study the manner in which $g(x)$ approaches $1/2$, let us write $g(x)=1/2+\epsilon$, where $\epsilon > 0$ is small. Then, \eqref{eq:g_n_alternative} gives (after dividing by the now-finite $e^{-x}$)
\begin{equation} \label{eq:g_eq_series_2}
    \sum_{n=0}^{\infty} \frac{(-\epsilon)^{(n)} x^n}{(1/2)^{(n)}n!}
    \approx 1 -\epsilon\sum_{n=1}^{\infty} \frac{x^n}{n(1/2)^{(n)}}
    = 0 \, ,
\end{equation}
where we kept only the leading $\epsilon$ contribution. The series can be written in terms of hypergeometric functions as
\begin{equation} \label{eq:series}
    \sum_{n=1}^{\infty} \frac{x^n}{n(1/2)^{(n)}} = 2x\sum_{n=0}^{\infty} \frac{1^{(n)} 1^{(n)} x^n}{(3/2)^{(n)}2^{(n)}n!} \equiv \tFt\qty(1,1;\frac{3}{2},2;x) \sim e^x \, .
\end{equation}
The equality is non-trivial, but can be shown to hold by a term-by-term comparison. The last approximation applies only roughly, but it indicates that the $\tFt$ function grows fast (quasi-exponentially) in x. Thus,
\begin{equation} \label{eq:g_for_large_x}
    g(x) = \frac{1}{2} + \frac{1}{2x \, \tFt\qty(1,1;\frac{3}{2},2;x)} \, , \quad x \gg 1
\end{equation}
approaches $1/2$ fast.

\begin{figure}
    \centering
    \includegraphics{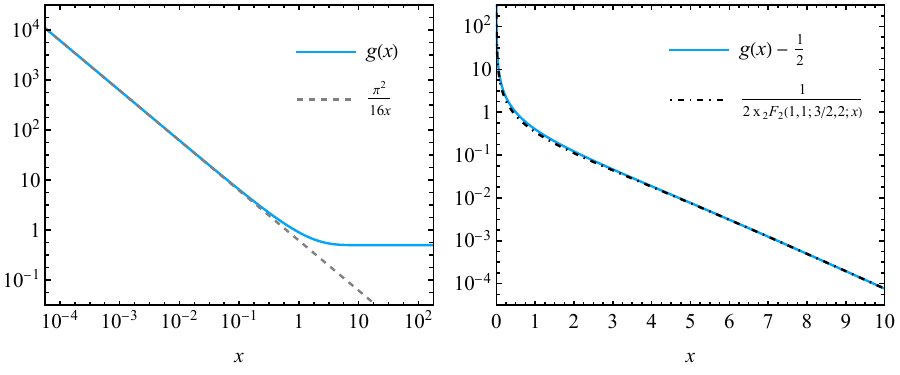}
    \caption{Asymptotic behaviour of the $g$ function and its apporximations.}
    \label{fig:g_approx}
\end{figure}

Figure~\ref{fig:g_approx} compares the asymptotic forms \eqref{eq:g_for_small_x} and \eqref{eq:g_for_large_x} to the numerically solved $g(x)$. The match is excellent in the asymptotic regions.

\section{Drift and diffusion during constant roll}
\label{sec:drift_and_diffusion_in_cr}

Let us discuss the evolution of the inflaton and its perturbations for the parabolic potential \eqref{eq:linear_drift_V}. We assume the Hubble parameter $H$ is approximately a constant and follow \cite{Karam:2022nym}.

In potential \eqref{eq:linear_drift_V}, in terms of e-folds $N$, the background inflaton follows the equation of motion
\begin{equation} \label{eq:phi_cr_eom}
    \partial_N^2\phi + 3\partial_N\phi + 3\eta_V\phi = 0
\end{equation}
with the solution\footnote{The constant $\nu$ was denoted by $\lambda$ in \cite{Karam:2022nym}; we changed the notation to avoid confusion with the Fokker--Planck eigenvalues.}
\begin{equation} \label{eq:phi_cr_solutions}
    \phi = \underbrace{c_-e^{\qty(-\frac{3}{2}-\nu)N}}_{\equiv\phi_-} + \underbrace{c_+e^{\qty(-\frac{3}{2}+\nu)N}}_{\equiv\phi_+} \, , \qquad \nu = \frac{3}{2}\sqrt{1-\frac{4}{3}\eta_V} \, ,
\end{equation}
where $c_-$ and $c_+$ are constants set by boundary conditions.
One of the solutions ($\phi_-$) dominates at early times, and the other ($\phi_+$) dominates at late times becoming an attractor, as long as $\eta_V < 3/4$ so that $\nu$ is real and positive. The exponents correspond to values of the second slow-roll parameter when the solution dominates (up to a minus sign):
\begin{equation} \label{eq:cr_etaH}
    \eta_H \xrightarrow{N \to -\infty} \eta_{H-} = \frac{3}{2}+\nu \, , \qquad
    \eta_H \xrightarrow{N \to +\infty} \eta_{H+} = \frac{3}{2}-\nu \, .
\end{equation}
During the stochastic evolution, we assume the field moves back and forth along the attractor $\phi_+$, with the constant $\eta_H=\eta_{H+}$. This is a good approximation due to the attractor behaviour and the squeezing of the perturbations that contribute the stochastic noise \cite{Tomberg:2022mkt}. Even if stochastic kicks moved the field outside of the attractor, it would quickly return there as the second solution $\phi_-$ decayed. With $\phi=\phi_+$, we can then write down the simplified classical equation of motion
\begin{equation} \label{eq:cr_attractor_eom}
    \partial_N \phi = -\eta_{H+}\phi \, .
\end{equation}
This is the drift $-\cV'$ used in  \eqref{eq:linear_drift_diffusion_drift_cr}, where we dropped the index `+' for simplicity. There is one subtlety: the attractor trajectory is restricted to either positive or negative $\phi$ (depending on the sign of $c_+$) and does not extend over $\phi=0$. If the stochastic motion takes the field over $\phi=0$, we assume the field quickly settles on the mirrored attractor on the other side, so that \eqref{eq:cr_attractor_eom} is justified also for such trajectories. For more details of the stochastic process, see \cite{Tomberg:2023kli}.

If $\eta > 3/4$, corresponding to a potential with a steep minimum, the $\nu$ in \eqref{eq:phi_cr_solutions} develops an imaginary part, corresponding to oscillating solutions. In the oscillating regime, the simple attractor behaviour is lost, and the analysis here becomes invalid. In this case, one needs to keep track of both the field and its momentum separately and apply stochastic kicks to both, using the full equation of motion \eqref{eq:phi_cr_eom} instead of the simpler attractor \eqref{eq:cr_attractor_eom}, as was done numerically in \cite{Figueroa:2020jkf, Figueroa:2021zah}. We don't pursue such a computation in this paper.

The field perturbations $\delta\phi$ in the spatially flat gauge can be written in terms of the Sasaki--Mukhanov variable $u$ as $u = a\delta\phi$. In Fourier space, to linear order, $u$ follows the equation of motion
\begin{equation} \label{eq:sasaki_mukhanov}
    u_k'' + \qty(k^2 - \frac{z''}{z})u_k = 0 \, , \qquad z \equiv a\partial_N \phi \, .
\end{equation}
Here prime denotes a derivative with respect to the conformal time $\tau=-1/(aH)$, where $a\sim e^N$ is the scale factor (note that $\tau \to -\infty$ at early times, and $\tau \to 0$ at late times).
In the background \eqref{eq:phi_cr_solutions}, we have
\begin{equation} \label{eq:cr_ddz_over_z}
    \frac{z''}{z} = \qty(\nu^2 - \frac{1}{4})\frac{1}{\tau^2} \, .
\end{equation}
Remarkably, this is a constant, independent of time and also of $c_-$ and $c_+$. In fact, \eqref{eq:cr_ddz_over_z} does not depend on the stochasticity of the background in any way, since the only time dependence is in $\tau\sim e^{-N}$ and $N$ is a non-stochastic clock variable (for a general discussion of the effects of the background stochasticity on the mode equations, see \cite{Tomberg:2024evi}). Thus, \eqref{eq:sasaki_mukhanov} can be solved self-consistently even in a stochastic background to yield
\begin{equation} \label{eq:u_solution}
    u_k(\tau) = \frac{\sqrt{-\tau\pi}}{2}H_\nu(-k\tau) \, ,
\end{equation}
where $H_\nu$ is the Hankel function of the first kind (not to be confused with the Hubble parameter $H$). The solution was chosen to satisfy the Bunch--Davies initial conditions \cite{Birrell:1982ix} in the sub-Hubble limit $-k\tau \gg 1$. At late times, in the super-Hubble limit $-k\tau \ll 1$, it gives (up to an irrelevant phase)
\begin{equation} \label{eq:u_solution_late_time}
    \delta\phi_k = \frac{u_k}{a} \xrightarrow{-k\tau \to 0} \frac{\sqrt{-\tau}}{2\sqrt{\pi}a}\qty(\frac{-k\tau}{2})^{-\nu}\Gamma(\nu) \, .
\end{equation}
For the purposes of stochastic inflation, the diffusion strength $\sigma^2$ in \eqref{eq:fokker_planck} is given by the power spectrum of $\delta\phi$ evaluated at the coarse-graining scale $k=k_\sigma = \sigma_c aH$ \cite{Tomberg:2024evi}. Here $\sigma_c < 1$ is a constant that sets the coarse-graining scale slightly above the Hubble radius, so that the coarse-grained field approximately follows the background equation \eqref{eq:phi_cr_eom}, plus stochastic corrections. Assuming the limit \eqref{eq:u_solution_late_time} applies at coarse-graining, we get\footnote{This result applies for modes that spend their whole `lifetime' from sub-Hubble to super-Hubble while the field is in the parabolic potential \eqref{eq:linear_drift_V}. If the parabolic section is only a part of a full potential, as is the case in PBH formation, then there is a transition period when the field enters the section, and modes that exit the Hubble radius around this period have a more complicated evolution. These transitionary modes form the peak of the curvature power spectrum and are important for PBH formation; they are not important for the long-time limit of eternal inflation, where \eqref{eq:u_solution_late_time} is sufficient.}
\begin{equation} \label{eq:cr_sigma}
    \sigma^2(\tau) = \Pphi(k_\sigma,\tau) = \frac{k_\sigma^3}{2\pi^2}|\delta\phi_{k_\sigma}(\tau)|^2 =
    \frac{H^2}{4\pi^2}\qty[\frac{\Gamma(\nu)}{\Gamma(3/2)}]^2 \qty(\frac{4}{\sigma^2_c})^{\nu-3/2} \, ,
\end{equation}
the result quoted in \eqref{eq:linear_drift_diffusion_drift_cr} in terms of $\eta_{H+}$.
Generally, the power spectrum depends on $\tau$ and $k$ and thus on $\sigma_c$, unless we take the de Sitter limit $\eta_V \to 0$, $\nu \to 3/2$ where it freezes to the well-known constant $H^2/(4\pi^2)$.\footnote{In contrast, the curvature perturbation $\R \sim \delta\phi/\sqrt{\epsilon_1}$ always (as long as the inflaton is on an attractor) freezes to a time-independent value at super-Hubble scales but may, of course, still depend on $k$.}${}^{,}$\footnote{The de Sitter limit is often used for the noise in stochastic inflation, see, e.g., \cite{Vennin:2020kng, Vennin:2024yzl}. Since we're going beyond the slow-roll approximation, we need to use the improved result \eqref{eq:cr_sigma}. This result is compatible with the constant-roll study of \cite{Tomberg:2023kli}, although in \cite{Tomberg:2023kli} the spectrum was presented through the frozen curvature perturbations.}
For moderate $\nu$ around this, the result \eqref{eq:cr_sigma} is of order $H$. However, if $\nu \to 0$, that is, $\eta_{H+}\to 3/2$, the problematic oscillating limit from above, $\sigma^2$ seems to diverge, as $\Gamma(\nu) \to \infty$. This is only an illusion: as $\nu$ decreases, the asymptotic behaviour \eqref{eq:u_solution_late_time} is reached later and later, and it may not apply at coarse-graining. Instead of \eqref{eq:cr_sigma}, one should use the full result
\begin{equation} \label{eq:cr_sigma_full}
    \sigma^2(\tau) = \frac{H^2}{4\pi^2}\frac{\sigma_c^3\pi}{2}\qty|H_\nu(\sigma_c)|^2 \, .
\end{equation}
This is perfectly regular in the $\nu \to 0$ limit: for example, for $\sigma_c=0.01$, we have $|H_\nu(\sigma_c)|^2 \approx 10.0$. Indeed, due to the $\sigma^3_c$ suppression, the diffusion strength can become extremely small in this limit compared to the de Sitter result.

Let us comment more on the $\sigma_c$ dependence of $\sigma$. The dependence may seem surprising, since $\sigma_c$ is chosen by hand and ideally should not affect the results -- yet it does. This is an inevitable consequence of perturbations whose spectrum is not scale invariant, and the effect may be important far from slow-roll. Physically, different choices of $\sigma_c$ correspond to background patches of different sizes, and in principle, the size may affect the patch's averaged evolution. A small $\sigma_c$ may also be required to allow the quantum fluctuations enough time to decohere into classical stochastic variables, an effect studied in the context of eternal inflation in \cite{Boddy:2016zkn}. However, the choice of $\sigma_c$ is also related to the goodness of the stochastic approximation; if the dependence on $\sigma_c$ is strong, one may question the validity of the approximations made. See \cite{Tomberg:2024evi} for a discussion about the choice of the coarse-graining scale in stochastic inflation.

\bibliographystyle{JHEP}
\bibliography{PBHeternal}

@article{Starobinsky:1980te,
    author = "Starobinsky, Alexei A.",
    editor = "Khalatnikov, I. M. and Mineev, V. P.",
    title = "{A New Type of Isotropic Cosmological Models Without Singularity}",
    doi = "10.1016/0370-2693(80)90670-X",
    journal = "Phys. Lett. B",
    volume = "91",
    pages = "99--102",
    year = "1980"
}

@article{Kazanas:1980tx,
    author = "Kazanas, D.",
    title = "{Dynamics of the Universe and Spontaneous Symmetry Breaking}",
    doi = "10.1086/183361",
    journal = "Astrophys. J. Lett.",
    volume = "241",
    pages = "L59--L63",
    year = "1980"
}

@article{Sato:1981qmu,
    author = "Sato, Katsuhiko",
    title = "{First-order phase transition of a vacuum and the expansion of the Universe}",
    doi = "10.1093/mnras/195.3.467",
    journal = "Mon. Not. Roy. Astron. Soc.",
    volume = "195",
    number = "3",
    pages = "467--479",
    year = "1981"
}

@article{Guth:1980zm,
    author = "Guth, Alan H.",
    editor = "Fang, Li-Zhi and Ruffini, R.",
    title = "{The Inflationary Universe: A Possible Solution to the Horizon and Flatness Problems}",
    reportNumber = "SLAC-PUB-2576",
    doi = "10.1103/PhysRevD.23.347",
    journal = "Phys. Rev. D",
    volume = "23",
    pages = "347--356",
    year = "1981"
}

@article{Carr:1974nx,
    author = "Carr, Bernard J. and Hawking, S. W.",
    title = "{Black holes in the early Universe}",
    doi = "10.1093/mnras/168.2.399",
    journal = "Mon. Not. Roy. Astron. Soc.",
    volume = "168",
    pages = "399--415",
    year = "1974"
}

@article{Carr:1975qj,
    author = "Carr, Bernard J.",
    title = "{The Primordial black hole mass spectrum}",
    doi = "10.1086/153853",
    journal = "Astrophys. J.",
    volume = "201",
    pages = "1--19",
    year = "1975"
}

@article{Green:2024bam,
    author = "Green, Anne M.",
    title = "{Primordial black holes as a dark matter candidate - a brief overview}",
    eprint = "2402.15211",
    archivePrefix = "arXiv",
    primaryClass = "astro-ph.CO",
    doi = "10.1016/j.nuclphysb.2024.116494",
    journal = "Nucl. Phys. B",
    volume = "1003",
    pages = "116494",
    year = "2024"
}

@inproceedings{Steinhardt:1982kg,
    author = "Steinhardt, Paul Joseph",
    title = "{NATURAL INFLATION}",
    booktitle = "{Nuffield Workshop on the Very Early Universe}",
    reportNumber = "UPR-0198T",
    month = "7",
    year = "1982"
}

@article{Vilenkin:1983xq,
    author = "Vilenkin, Alexander",
    title = "{The Birth of Inflationary Universes}",
    reportNumber = "TUTP-83-1",
    doi = "10.1103/PhysRevD.27.2848",
    journal = "Phys. Rev. D",
    volume = "27",
    pages = "2848",
    year = "1983"
}

@article{Linde:1986fc,
    author = "Linde, Andrei D.",
    title = "{ETERNAL CHAOTIC INFLATION}",
    reportNumber = "Print-86-0417 (LEBEDEV INST), IC/86/74",
    doi = "10.1142/S0217732386000129",
    journal = "Mod. Phys. Lett. A",
    volume = "1",
    pages = "81",
    year = "1986"
}

@article{Linde:1986fd,
    author = "Linde, Andrei D.",
    title = "{Eternally Existing Selfreproducing Chaotic Inflationary Universe}",
    reportNumber = "Print-86-0418 (LEBEDEV INST), LEBEDEV-86-106",
    doi = "10.1016/0370-2693(86)90611-8",
    journal = "Phys. Lett. B",
    volume = "175",
    pages = "395--400",
    year = "1986"
}

@article{Vilenkin:1999pi,
    author = "Vilenkin, Alexander",
    editor = "Kursunoglu, B. N. and Mintz, S. L. and Perlmutter, A.",
    title = "{Eternal inflation and the present universe}",
    eprint = "gr-qc/9911087",
    archivePrefix = "arXiv",
    doi = "10.1016/S0920-5632(00)00755-6",
    journal = "Nucl. Phys. B Proc. Suppl.",
    volume = "88",
    pages = "67--74",
    year = "2000"
}

@article{Vilenkin:2004vx,
    author = "Vilenkin, Alexander",
    title = "{Eternal inflation and chaotic terminology}",
    eprint = "gr-qc/0409055",
    archivePrefix = "arXiv",
    month = "9",
    year = "2004"
}

@article{Guth:2007ng,
    author = "Guth, Alan H.",
    editor = "Sola, Joan",
    title = "{Eternal inflation and its implications}",
    eprint = "hep-th/0702178",
    archivePrefix = "arXiv",
    reportNumber = "MIT-CTP-3811",
    doi = "10.1088/1751-8113/40/25/S25",
    journal = "J. Phys. A",
    volume = "40",
    pages = "6811--6826",
    year = "2007"
}

@book{Winitzki:2008zz,
    author = "Winitzki, Sergei",
    title = "{Eternal inflation}",
    doi = "10.1142/6923",
    year = "2008"
}

@article{Linde:2015edk,
    author = "Linde, Andrei",
    title = "{A brief history of the multiverse}",
    eprint = "1512.01203",
    archivePrefix = "arXiv",
    primaryClass = "hep-th",
    doi = "10.1088/1361-6633/aa50e4",
    journal = "Rept. Prog. Phys.",
    volume = "80",
    number = "2",
    pages = "022001",
    year = "2017"
}

@article{Pattison:2017mbe,
    author = "Pattison, Chris and Vennin, Vincent and Assadullahi, Hooshyar and Wands, David",
    title = "{Quantum diffusion during inflation and primordial black holes}",
    eprint = "1707.00537",
    archivePrefix = "arXiv",
    primaryClass = "hep-th",
    doi = "10.1088/1475-7516/2017/10/046",
    journal = "JCAP",
    volume = "10",
    pages = "046",
    year = "2017"
}

@article{Ezquiaga:2019ftu,
    author = "Ezquiaga, Jose Mar\'\i{}a and Garc\'\i{}a-Bellido, Juan and Vennin, Vincent",
    title = "{The exponential tail of inflationary fluctuations: consequences for primordial black holes}",
    eprint = "1912.05399",
    archivePrefix = "arXiv",
    primaryClass = "astro-ph.CO",
    doi = "10.1088/1475-7516/2020/03/029",
    journal = "JCAP",
    volume = "03",
    pages = "029",
    year = "2020"
}

@article{Pattison:2021oen,
    author = "Pattison, Chris and Vennin, Vincent and Wands, David and Assadullahi, Hooshyar",
    title = "{Ultra-slow-roll inflation with quantum diffusion}",
    eprint = "2101.05741",
    archivePrefix = "arXiv",
    primaryClass = "astro-ph.CO",
    doi = "10.1088/1475-7516/2021/04/080",
    journal = "JCAP",
    volume = "04",
    pages = "080",
    year = "2021"
}

@phdthesis{Vennin:2020kng,
    author = "Vennin, Vincent",
    title = "{Stochastic inflation and primordial black holes}",
    eprint = "2009.08715",
    archivePrefix = "arXiv",
    primaryClass = "astro-ph.CO",
    school = "U. Paris-Saclay",
    month = "6",
    year = "2020"
}

@inbook{Vennin:2024yzl,
    author = "Vennin, Vincent and Wands, David",
    title = "{Quantum Diffusion and~Large Primordial Perturbations from~Inflation}",
    eprint = "2402.12672",
    archivePrefix = "arXiv",
    primaryClass = "astro-ph.CO",
    doi = "10.1007/978-981-97-8887-3_8",
    year = "2025"
}

@article{Tomberg:2024evi,
    author = "Tomberg, Eemeli",
    title = "{It\^o, Stratonovich, and zoom-in schemes in stochastic inflation}",
    eprint = "2411.12465",
    archivePrefix = "arXiv",
    primaryClass = "astro-ph.CO",
    doi = "10.1088/1475-7516/2025/04/035",
    journal = "JCAP",
    volume = "04",
    pages = "035",
    year = "2025"
}

@article{Linde:1993xx,
    author = "Linde, Andrei D. and Linde, Dmitri A. and Mezhlumian, Arthur",
    title = "{From the Big Bang theory to the theory of a stationary universe}",
    eprint = "gr-qc/9306035",
    archivePrefix = "arXiv",
    reportNumber = "SU-ITP-93-13",
    doi = "10.1103/PhysRevD.49.1783",
    journal = "Phys. Rev. D",
    volume = "49",
    pages = "1783--1826",
    year = "1994"
}

@article{Garcia-Bellido:1994gng,
    author = "Garcia-Bellido, Juan and Linde, Andrei D.",
    title = "{Stationarity of inflation and predictions of quantum cosmology}",
    eprint = "hep-th/9408023",
    archivePrefix = "arXiv",
    reportNumber = "SU-ITP-94-24, IEM-FT-88-94",
    doi = "10.1103/PhysRevD.51.429",
    journal = "Phys. Rev. D",
    volume = "51",
    pages = "429--443",
    year = "1995"
}

@article{Winitzki:2001np,
    author = "Winitzki, Serge",
    title = "{The Eternal fractal in the universe}",
    eprint = "gr-qc/0111048",
    archivePrefix = "arXiv",
    doi = "10.1103/PhysRevD.65.083506",
    journal = "Phys. Rev. D",
    volume = "65",
    pages = "083506",
    year = "2002"
}

@article{Winitzki:2005ya,
    author = "Winitzki, Sergei",
    title = "{On time-reparametrization invariance in eternal inflation}",
    eprint = "gr-qc/0504084",
    archivePrefix = "arXiv",
    doi = "10.1103/PhysRevD.71.123507",
    journal = "Phys. Rev. D",
    volume = "71",
    pages = "123507",
    year = "2005"
}

@article{Linde:1993nz,
    author = "Linde, Andrei D. and Mezhlumian, Arthur",
    title = "{Stationary universe}",
    eprint = "gr-qc/9304015",
    archivePrefix = "arXiv",
    reportNumber = "SU-ITP-93-9",
    doi = "10.1016/0370-2693(93)90187-M",
    journal = "Phys. Lett. B",
    volume = "307",
    pages = "25--33",
    year = "1993"
}

@article{Rudelius:2019cfh,
    author = "Rudelius, Tom",
    title = "{Conditions for (No) Eternal Inflation}",
    eprint = "1905.05198",
    archivePrefix = "arXiv",
    primaryClass = "hep-th",
    doi = "10.1088/1475-7516/2019/08/009",
    journal = "JCAP",
    volume = "08",
    pages = "009",
    year = "2019"
}

@article{Creminelli:2008es,
    author = "Creminelli, Paolo and Dubovsky, Sergei and Nicolis, Alberto and Senatore, Leonardo and Zaldarriaga, Matias",
    title = "{The Phase Transition to Slow-roll Eternal Inflation}",
    eprint = "0802.1067",
    archivePrefix = "arXiv",
    primaryClass = "hep-th",
    reportNumber = "IC-2008-002",
    doi = "10.1088/1126-6708/2008/09/036",
    journal = "JHEP",
    volume = "09",
    pages = "036",
    year = "2008"
}

@article{Tegmark:2004qd,
    author = "Tegmark, Max",
    title = "{What does inflation really predict?}",
    eprint = "astro-ph/0410281",
    archivePrefix = "arXiv",
    doi = "10.1088/1475-7516/2005/04/001",
    journal = "JCAP",
    volume = "04",
    pages = "001",
    year = "2005"
}

@article{Greenwood:2021uuj,
    author = "Greenwood, Ross N. and Aguirre, Anthony",
    title = "{How generic is eternal inflation?}",
    eprint = "2111.14218",
    archivePrefix = "arXiv",
    primaryClass = "gr-qc",
    month = "11",
    year = "2021"
}

@article{Planck:2018jri,
    author = "Akrami, Y. and others",
    collaboration = "Planck",
    title = "{Planck 2018 results. X. Constraints on inflation}",
    eprint = "1807.06211",
    archivePrefix = "arXiv",
    primaryClass = "astro-ph.CO",
    doi = "10.1051/0004-6361/201833887",
    journal = "Astron. Astrophys.",
    volume = "641",
    pages = "A10",
    year = "2020"
}

@article{ACT:2025fju,
    author = "Louis, Thibaut and others",
    collaboration = "ACT",
    title = "{The Atacama Cosmology Telescope: DR6 Power Spectra, Likelihoods and $\Lambda$CDM Parameters}",
    eprint = "2503.14452",
    archivePrefix = "arXiv",
    primaryClass = "astro-ph.CO",
    reportNumber = "FERMILAB-PUB-25-0071-PPD",
    month = "3",
    year = "2025"
}

@article{Barenboim:2016mmw,
    author = "Barenboim, Gabriela and Park, Wan-Il and Kinney, William H.",
    title = "{Eternal Hilltop Inflation}",
    eprint = "1601.08140",
    archivePrefix = "arXiv",
    primaryClass = "astro-ph.CO",
    reportNumber = "FTUV-16-01-29, IFIC-16-03",
    doi = "10.1088/1475-7516/2016/05/030",
    journal = "JCAP",
    volume = "05",
    pages = "030",
    year = "2016"
}

@article{Motohashi:2014ppa,
    author = "Motohashi, Hayato and Starobinsky, Alexei A. and Yokoyama, Jun'ichi",
    title = "{Inflation with a constant rate of roll}",
    eprint = "1411.5021",
    archivePrefix = "arXiv",
    primaryClass = "astro-ph.CO",
    reportNumber = "RESCEU-51-14",
    doi = "10.1088/1475-7516/2015/09/018",
    journal = "JCAP",
    volume = "09",
    pages = "018",
    year = "2015"
}

@article{Finelli:2008zg,
    author = "Finelli, F. and Marozzi, G. and Starobinsky, A. A. and Vacca, G. P. and Venturi, G.",
    title = "{Generation of fluctuations during inflation: Comparison of stochastic and field-theoretic approaches}",
    eprint = "0808.1786",
    archivePrefix = "arXiv",
    primaryClass = "hep-th",
    doi = "10.1103/PhysRevD.79.044007",
    journal = "Phys. Rev. D",
    volume = "79",
    pages = "044007",
    year = "2009"
}

@article{Briaud:2023eae,
    author = "Briaud, Vadim and Vennin, Vincent",
    title = "{Uphill inflation}",
    eprint = "2301.09336",
    archivePrefix = "arXiv",
    primaryClass = "astro-ph.CO",
    doi = "10.1088/1475-7516/2023/06/029",
    journal = "JCAP",
    volume = "06",
    pages = "029",
    year = "2023"
}

@article{BICEP:2021xfz,
    author = "Ade, P. A. R. and others",
    collaboration = "BICEP, Keck",
    title = "{Improved Constraints on Primordial Gravitational Waves using Planck, WMAP, and BICEP/Keck Observations through the 2018 Observing Season}",
    eprint = "2110.00483",
    archivePrefix = "arXiv",
    primaryClass = "astro-ph.CO",
    doi = "10.1103/PhysRevLett.127.151301",
    journal = "Phys. Rev. Lett.",
    volume = "127",
    number = "15",
    pages = "151301",
    year = "2021"
}

@article{Karam:2022nym,
    author = {Karam, Alexandros and Koivunen, Niko and Tomberg, Eemeli and Vaskonen, Ville and Veerm\"ae, Hardi},
    title = "{Anatomy of single-field inflationary models for primordial black holes}",
    eprint = "2205.13540",
    archivePrefix = "arXiv",
    primaryClass = "astro-ph.CO",
    doi = "10.1088/1475-7516/2023/03/013",
    journal = "JCAP",
    volume = "03",
    pages = "013",
    year = "2023"
}

@article{Tomberg:2023kli,
    author = "Tomberg, Eemeli",
    title = "{Stochastic constant-roll inflation and primordial black holes}",
    eprint = "2304.10903",
    archivePrefix = "arXiv",
    primaryClass = "astro-ph.CO",
    doi = "10.1103/PhysRevD.108.043502",
    journal = "Phys. Rev. D",
    volume = "108",
    number = "4",
    pages = "043502",
    year = "2023"
}

@article{Kannike:2017bxn,
    author = {Kannike, Kristjan and Marzola, Luca and Raidal, Martti and Veerm\"ae, Hardi},
    title = "{Single Field Double Inflation and Primordial Black Holes}",
    eprint = "1705.06225",
    archivePrefix = "arXiv",
    primaryClass = "astro-ph.CO",
    doi = "10.1088/1475-7516/2017/09/020",
    journal = "JCAP",
    volume = "09",
    pages = "020",
    year = "2017"
}

@article{Ballesteros:2017fsr,
    author = "Ballesteros, Guillermo and Taoso, Marco",
    title = "{Primordial black hole dark matter from single field inflation}",
    eprint = "1709.05565",
    archivePrefix = "arXiv",
    primaryClass = "hep-ph",
    doi = "10.1103/PhysRevD.97.023501",
    journal = "Phys. Rev. D",
    volume = "97",
    number = "2",
    pages = "023501",
    year = "2018"
}

@article{Dalianis:2018frf,
    author = "Dalianis, Ioannis and Kehagias, Alex and Tringas, George",
    title = "{Primordial black holes from \ensuremath{\alpha}-attractors}",
    eprint = "1805.09483",
    archivePrefix = "arXiv",
    primaryClass = "astro-ph.CO",
    doi = "10.1088/1475-7516/2019/01/037",
    journal = "JCAP",
    volume = "01",
    pages = "037",
    year = "2019"
}

@article{Mishra:2019pzq,
    author = "Mishra, Swagat S. and Sahni, Varun",
    title = "{Primordial Black Holes from a tiny bump/dip in the Inflaton potential}",
    eprint = "1911.00057",
    archivePrefix = "arXiv",
    primaryClass = "gr-qc",
    doi = "10.1088/1475-7516/2020/04/007",
    journal = "JCAP",
    volume = "04",
    pages = "007",
    year = "2020"
}

@article{Ballesteros:2020qam,
    author = "Ballesteros, Guillermo and Rey, Juli\'an and Taoso, Marco and Urbano, Alfredo",
    title = "{Primordial black holes as dark matter and gravitational waves from single-field polynomial inflation}",
    eprint = "2001.08220",
    archivePrefix = "arXiv",
    primaryClass = "astro-ph.CO",
    doi = "10.1088/1475-7516/2020/07/025",
    journal = "JCAP",
    volume = "07",
    pages = "025",
    year = "2020"
}

@article{Hawking:1971ei,
    author = "Hawking, Stephen",
    title = "{Gravitationally collapsed objects of very low mass}",
    doi = "10.1093/mnras/152.1.75",
    journal = "Mon. Not. Roy. Astron. Soc.",
    volume = "152",
    pages = "75",
    year = "1971"
}

@article{Green:2020jor,
    author = "Green, Anne M. and Kavanagh, Bradley J.",
    title = "{Primordial Black Holes as a dark matter candidate}",
    eprint = "2007.10722",
    archivePrefix = "arXiv",
    primaryClass = "astro-ph.CO",
    doi = "10.1088/1361-6471/abc534",
    journal = "J. Phys. G",
    volume = "48",
    number = "4",
    pages = "043001",
    year = "2021"
}

@article{Carr:2025kdk,
    author = "Carr, Bernard and Kuhnel, Florian",
    title = "{Primordial Black Holes}",
    eprint = "2502.15279",
    archivePrefix = "arXiv",
    primaryClass = "astro-ph.CO",
    month = "2",
    year = "2025"
}

@article{Sato:1981bf,
    author = "Sato, Katsuhiko and Sasaki, Misao and Kodama, Hideo and Maeda, Kei-ichi",
    title = "{Creation of Wormholes by First Order Phase Transition of a Vacuum in the Early Universe}",
    reportNumber = "KUNS 574",
    doi = "10.1143/PTP.65.1443",
    journal = "Prog. Theor. Phys.",
    volume = "65",
    pages = "1443",
    year = "1981"
}

@article{Sato:1981gv,
    author = "Sato, Katsuhiko and Kodama, Hideo and Sasaki, Misao and Maeda, Kei-ichi",
    title = "{Multiproduction of Universes by First Order Phase Transition of a Vacuum}",
    reportNumber = "KUNS 588",
    doi = "10.1016/0370-2693(82)91152-2",
    journal = "Phys. Lett. B",
    volume = "108",
    pages = "103--107",
    year = "1982"
}

@article{Blau:1986cw,
    author = "Blau, Steven K. and Guendelman, E. I. and Guth, Alan H.",
    title = "{The Dynamics of False Vacuum Bubbles}",
    reportNumber = "MIT-CTP-1292",
    doi = "10.1103/PhysRevD.35.1747",
    journal = "Phys. Rev. D",
    volume = "35",
    pages = "1747",
    year = "1987"
}

@article{Aryal:1987vn,
    author = "Aryal, Mukunda and Vilenkin, Alexander",
    title = "{The Fractal Dimension of Inflationary Universe}",
    reportNumber = "TUTP-87-11",
    doi = "10.1016/0370-2693(87)90932-4",
    journal = "Phys. Lett. B",
    volume = "199",
    pages = "351--357",
    year = "1987"
}

@article{Neeman:1994fxj,
    author = "Ne'eman, Yuval",
    editor = "Ruffini, R. and Verbin, Y.",
    title = "{Cosmological surrealism: More than 'eternal reality' is needed}",
    eprint = "hep-th/9403087",
    archivePrefix = "arXiv",
    reportNumber = "TAUP-N237-94",
    doi = "10.1007/BF02189251",
    journal = "Found. Phys. Lett.",
    volume = "7",
    pages = "483--488",
    year = "1994"
}

@article{Bousso:2006ge,
    author = "Bousso, Raphael and Freivogel, Ben and Yang, I-Sheng",
    title = "{Eternal Inflation: The Inside Story}",
    eprint = "hep-th/0606114",
    archivePrefix = "arXiv",
    reportNumber = "UCB-PTH-06-09, LBNL-60250",
    doi = "10.1103/PhysRevD.74.103516",
    journal = "Phys. Rev. D",
    volume = "74",
    pages = "103516",
    year = "2006"
}

@article{Kopp:2010sh,
    author = "Kopp, Michael and Hofmann, Stefan and Weller, Jochen",
    title = "{Separate Universes Do Not Constrain Primordial Black Hole Formation}",
    eprint = "1012.4369",
    archivePrefix = "arXiv",
    primaryClass = "astro-ph.CO",
    doi = "10.1103/PhysRevD.83.124025",
    journal = "Phys. Rev. D",
    volume = "83",
    pages = "124025",
    year = "2011"
}

@article{Carr:2014pga,
    author = "Carr, B. J. and Harada, Tomohiro",
    title = "{Separate universe problem: 40 years on}",
    eprint = "1405.3624",
    archivePrefix = "arXiv",
    primaryClass = "astro-ph.CO",
    reportNumber = "RUP-14-8",
    doi = "10.1103/PhysRevD.91.084048",
    journal = "Phys. Rev. D",
    volume = "91",
    number = "8",
    pages = "084048",
    year = "2015"
}

@article{Escriva:2023uko,
    author = "Escriv\`a, Albert and Atal, Vicente and Garriga, Jaume",
    title = "{Formation of trapped vacuum bubbles during inflation, and consequences for PBH scenarios}",
    eprint = "2306.09990",
    archivePrefix = "arXiv",
    primaryClass = "astro-ph.CO",
    doi = "10.1088/1475-7516/2023/10/035",
    journal = "JCAP",
    volume = "10",
    pages = "035",
    year = "2023"
}

@article{Harada:2024jxl,
    author = "Harada, Tomohiro",
    title = "{Primordial Black Holes: Formation, Spin and Type II}",
    eprint = "2409.01934",
    archivePrefix = "arXiv",
    primaryClass = "gr-qc",
    reportNumber = "RUP-24-16",
    doi = "10.3390/universe10120444",
    journal = "Universe",
    volume = "10",
    number = "12",
    pages = "444",
    year = "2024"
}

@article{Uehara:2024yyp,
    author = "Uehara, Koichiro and Escriv\`a, Albert and Harada, Tomohiro and Saito, Daiki and Yoo, Chul-Moon",
    title = "{Numerical simulation of type II primordial black hole formation}",
    eprint = "2401.06329",
    archivePrefix = "arXiv",
    primaryClass = "gr-qc",
    reportNumber = "RUP-24-1",
    doi = "10.1088/1475-7516/2025/01/003",
    journal = "JCAP",
    volume = "01",
    pages = "003",
    year = "2025"
}

@article{Escriva:2025eqc,
    author = "Escriv\`a, Albert",
    title = "{A new approach for simulating PBH formation from generic curvature fluctuations with the Misner-Sharp formalism}",
    eprint = "2504.05813",
    archivePrefix = "arXiv",
    primaryClass = "astro-ph.CO",
    month = "4",
    year = "2025"
}

@article{Escriva:2025rja,
    author = "Escriv\`a, Albert",
    title = "{The threshold for PBH formation in the type-II region and its analytical estimation}",
    eprint = "2504.05814",
    archivePrefix = "arXiv",
    primaryClass = "astro-ph.CO",
    month = "4",
    year = "2025"
}

@article{Uehara:2025idq,
    author = "Uehara, Koichiro and Escriv\`a, Albert and Harada, Tomohiro and Saito, Daiki and Yoo, Chul-Moon",
    title = "{Primordial black hole formation from a type II perturbation in the absence and presence of pressure}",
    eprint = "2505.00366",
    archivePrefix = "arXiv",
    primaryClass = "gr-qc",
    reportNumber = "RUP-25-12",
    month = "5",
    year = "2025"
}

@article{Shimada:2024eec,
    author = "Shimada, Masaaki and Escriv\'a, Albert and Saito, Daiki and Uehara, Koichiro and Yoo, Chul-Moon",
    title = "{Primordial black hole formation from type II fluctuations with primordial non-Gaussianity}",
    eprint = "2411.07648",
    archivePrefix = "arXiv",
    primaryClass = "gr-qc",
    doi = "10.1088/1475-7516/2025/02/018",
    journal = "JCAP",
    volume = "02",
    pages = "018",
    year = "2025"
}

@article{Garriga:2015fdk,
    author = "Garriga, Jaume and Vilenkin, Alexander and Zhang, Jun",
    title = "{Black holes and the multiverse}",
    eprint = "1512.01819",
    archivePrefix = "arXiv",
    primaryClass = "hep-th",
    doi = "10.1088/1475-7516/2016/02/064",
    journal = "JCAP",
    volume = "02",
    pages = "064",
    year = "2016"
}

@article{Sekino:2010vc,
    author = "Sekino, Yasuhiro and Shenker, Stephen and Susskind, Leonard",
    title = "{On the Topological Phases of Eternal Inflation}",
    eprint = "1003.1347",
    archivePrefix = "arXiv",
    primaryClass = "hep-th",
    reportNumber = "SITP-10-04, OIQP-10-01",
    doi = "10.1103/PhysRevD.81.123515",
    journal = "Phys. Rev. D",
    volume = "81",
    pages = "123515",
    year = "2010"
}

@article{Garriga:2006hw,
    author = "Garriga, Jaume and Guth, Alan H. and Vilenkin, Alexander",
    title = "{Eternal inflation, bubble collisions, and the persistence of memory}",
    eprint = "hep-th/0612242",
    archivePrefix = "arXiv",
    reportNumber = "MIT-CTP-3800",
    doi = "10.1103/PhysRevD.76.123512",
    journal = "Phys. Rev. D",
    volume = "76",
    pages = "123512",
    year = "2007"
}

@article{Aguirre:2007an,
    author = "Aguirre, Anthony and Johnson, Matthew C and Shomer, Assaf",
    title = "{Towards observable signatures of other bubble universes}",
    eprint = "0704.3473",
    archivePrefix = "arXiv",
    primaryClass = "hep-th",
    doi = "10.1103/PhysRevD.76.063509",
    journal = "Phys. Rev. D",
    volume = "76",
    pages = "063509",
    year = "2007"
}

@article{Aguirre:2009ug,
    author = "Aguirre, Anthony and Johnson, Matthew C.",
    title = "{A Status report on the observability of cosmic bubble collisions}",
    eprint = "0908.4105",
    archivePrefix = "arXiv",
    primaryClass = "hep-th",
    doi = "10.1088/0034-4885/74/7/074901",
    journal = "Rept. Prog. Phys.",
    volume = "74",
    pages = "074901",
    year = "2011"
}

@article{Feeney:2010jj,
    author = "Feeney, Stephen M. and Johnson, Matthew C. and Mortlock, Daniel J. and Peiris, Hiranya V.",
    title = "{First Observational Tests of Eternal Inflation}",
    eprint = "1012.1995",
    archivePrefix = "arXiv",
    primaryClass = "astro-ph.CO",
    doi = "10.1103/PhysRevLett.107.071301",
    journal = "Phys. Rev. Lett.",
    volume = "107",
    pages = "071301",
    year = "2011"
}

@article{Kleban:2011pg,
    author = "Kleban, Matthew",
    title = "{Cosmic Bubble Collisions}",
    eprint = "1107.2593",
    archivePrefix = "arXiv",
    primaryClass = "astro-ph.CO",
    doi = "10.1088/0264-9381/28/20/204008",
    journal = "Class. Quant. Grav.",
    volume = "28",
    pages = "204008",
    year = "2011"
}

@article{Jain:2019gsq,
    author = "Jain, Mudit and Hertzberg, Mark P.",
    title = "{Statistics of Inflating Regions in Eternal Inflation}",
    eprint = "1904.04262",
    archivePrefix = "arXiv",
    primaryClass = "astro-ph.CO",
    doi = "10.1103/PhysRevD.100.023513",
    journal = "Phys. Rev. D",
    volume = "100",
    number = "2",
    pages = "023513",
    year = "2019"
}

@article{Linde:1994wt,
    author = "Linde, Andrei D. and Linde, Dmitri A.",
    title = "{Topological defects as seeds for eternal inflation}",
    eprint = "hep-th/9402115",
    archivePrefix = "arXiv",
    reportNumber = "SU-ITP-94-3",
    doi = "10.1103/PhysRevD.50.2456",
    journal = "Phys. Rev. D",
    volume = "50",
    pages = "2456--2468",
    year = "1994"
}

@article{Winitzki:2006rn,
    author = "Winitzki, Sergei",
    title = "{Predictions in eternal inflation}",
    eprint = "gr-qc/0612164",
    archivePrefix = "arXiv",
    doi = "10.1007/978-3-540-74353-8_5",
    journal = "Lect. Notes Phys.",
    volume = "738",
    pages = "157--191",
    year = "2008"
}

@article{Susskind:2003kw,
    author = "Susskind, Leonard",
    editor = "Carr, Bernard J.",
    title = "{The Anthropic landscape of string theory}",
    eprint = "hep-th/0302219",
    archivePrefix = "arXiv",
    pages = "247--266",
    month = "2",
    year = "2003"
}

@article{Linde:1994gy,
    author = "Linde, Andrei D. and Linde, Dmitri A. and Mezhlumian, Arthur",
    title = "{Do we live in the center of the world?}",
    eprint = "hep-th/9411111",
    archivePrefix = "arXiv",
    reportNumber = "SU-ITP-94-39",
    doi = "10.1016/0370-2693(94)01641-O",
    journal = "Phys. Lett. B",
    volume = "345",
    pages = "203--210",
    year = "1995"
}

@article{Freivogel:2011eg,
    author = "Freivogel, Ben",
    title = "{Making predictions in the multiverse}",
    eprint = "1105.0244",
    archivePrefix = "arXiv",
    primaryClass = "hep-th",
    doi = "10.1088/0264-9381/28/20/204007",
    journal = "Class. Quant. Grav.",
    volume = "28",
    pages = "204007",
    year = "2011"
}

@article{Garriga:2005av,
    author = "Garriga, Jaume and Schwartz-Perlov, Delia and Vilenkin, Alexander and Winitzki, Sergei",
    title = "{Probabilities in the inflationary multiverse}",
    eprint = "hep-th/0509184",
    archivePrefix = "arXiv",
    doi = "10.1088/1475-7516/2006/01/017",
    journal = "JCAP",
    volume = "01",
    pages = "017",
    year = "2006"
}

@article{Feldstein:2005bm,
    author = "Feldstein, Brian and Hall, Lawrence J. and Watari, Taizan",
    title = "{Density perturbations and the cosmological constant from inflationary landscapes}",
    eprint = "hep-th/0506235",
    archivePrefix = "arXiv",
    reportNumber = "UCB-PTH-05-17, LBNL-57714",
    doi = "10.1103/PhysRevD.72.123506",
    journal = "Phys. Rev. D",
    volume = "72",
    pages = "123506",
    year = "2005"
}

@article{Panagopoulos:2019ail,
    author = "Panagopoulos, George and Silverstein, Eva",
    title = "{Primordial Black Holes from non-Gaussian tails}",
    eprint = "1906.02827",
    archivePrefix = "arXiv",
    primaryClass = "hep-th",
    month = "6",
    year = "2019"
}

@article{Figueroa:2020jkf,
    author = "Figueroa, Daniel G. and Raatikainen, Sami and Rasanen, Syksy and Tomberg, Eemeli",
    title = "{Non-Gaussian Tail of the Curvature Perturbation in Stochastic Ultraslow-Roll Inflation: Implications for Primordial Black Hole Production}",
    eprint = "2012.06551",
    archivePrefix = "arXiv",
    primaryClass = "astro-ph.CO",
    reportNumber = "HIP-2020-32/TH",
    doi = "10.1103/PhysRevLett.127.101302",
    journal = "Phys. Rev. Lett.",
    volume = "127",
    number = "10",
    pages = "101302",
    year = "2021"
}

@article{Biagetti:2021eep,
    author = "Biagetti, Matteo and De Luca, Valerio and Franciolini, Gabriele and Kehagias, Alex and Riotto, Antonio",
    title = "{The formation probability of primordial black holes}",
    eprint = "2105.07810",
    archivePrefix = "arXiv",
    primaryClass = "astro-ph.CO",
    doi = "10.1016/j.physletb.2021.136602",
    journal = "Phys. Lett. B",
    volume = "820",
    pages = "136602",
    year = "2021"
}

@article{Kitajima:2021fpq,
    author = "Kitajima, Naoya and Tada, Yuichiro and Yokoyama, Shuichiro and Yoo, Chul-Moon",
    title = "{Primordial black holes in peak theory with a non-Gaussian tail}",
    eprint = "2109.00791",
    archivePrefix = "arXiv",
    primaryClass = "astro-ph.CO",
    reportNumber = "TU-1130",
    doi = "10.1088/1475-7516/2021/10/053",
    journal = "JCAP",
    volume = "10",
    pages = "053",
    year = "2021"
}

@article{Figueroa:2021zah,
    author = "Figueroa, Daniel G. and Raatikainen, Sami and Rasanen, Syksy and Tomberg, Eemeli",
    title = "{Implications of stochastic effects for primordial black hole production in ultra-slow-roll inflation}",
    eprint = "2111.07437",
    archivePrefix = "arXiv",
    primaryClass = "astro-ph.CO",
    reportNumber = "HIP-2021-31/TH",
    doi = "10.1088/1475-7516/2022/05/027",
    journal = "JCAP",
    volume = "05",
    number = "05",
    pages = "027",
    year = "2022"
}

@article{Tomberg:2021xxv,
    author = "Tomberg, Eemeli",
    title = "{A numerical approach to stochastic inflation and primordial black holes}",
    eprint = "2110.10684",
    archivePrefix = "arXiv",
    primaryClass = "astro-ph.CO",
    doi = "10.1088/1742-6596/2156/1/012010",
    journal = "J. Phys. Conf. Ser.",
    volume = "2156",
    number = "1",
    pages = "012010",
    year = "2021"
}

@article{Hooshangi:2021ubn,
    author = "Hooshangi, Sina and Namjoo, Mohammad Hossein and Noorbala, Mahdiyar",
    title = "{Rare events are nonperturbative: Primordial black holes from heavy-tailed distributions}",
    eprint = "2112.04520",
    archivePrefix = "arXiv",
    primaryClass = "astro-ph.CO",
    doi = "10.1016/j.physletb.2022.137400",
    journal = "Phys. Lett. B",
    volume = "834",
    pages = "137400",
    year = "2022"
}

@article{Cai:2021zsp,
    author = "Cai, Yi-Fu and Ma, Xiao-Han and Sasaki, Misao and Wang, Dong-Gang and Zhou, Zihan",
    title = "{One small step for an inflaton, one giant leap for inflation: A novel non-Gaussian tail and primordial black holes}",
    eprint = "2112.13836",
    archivePrefix = "arXiv",
    primaryClass = "astro-ph.CO",
    reportNumber = "YITP-22-143, YITP-21-143",
    doi = "10.1016/j.physletb.2022.137461",
    journal = "Phys. Lett. B",
    volume = "834",
    pages = "137461",
    year = "2022"
}

@article{Achucarro:2021pdh,
    author = "Achucarro, Ana and Cespedes, Sebastian and Davis, Anne-Christine and Palma, Gonzalo A.",
    title = "{The hand-made tail: non-perturbative tails from multifield inflation}",
    eprint = "2112.14712",
    archivePrefix = "arXiv",
    primaryClass = "hep-th",
    doi = "10.1007/JHEP05(2022)052",
    journal = "JHEP",
    volume = "05",
    pages = "052",
    year = "2022"
}

@article{Tomberg:2022mkt,
    author = "Tomberg, Eemeli",
    title = "{Numerical stochastic inflation constrained by frozen noise}",
    eprint = "2210.17441",
    archivePrefix = "arXiv",
    primaryClass = "astro-ph.CO",
    doi = "10.1088/1475-7516/2023/04/042",
    journal = "JCAP",
    volume = "04",
    pages = "042",
    year = "2023"
}

@article{DeLuca:2022rfz,
    author = "De Luca, V. and Riotto, A.",
    title = "{A note on the abundance of primordial black holes: Use and misuse of the metric curvature perturbation}",
    eprint = "2201.09008",
    archivePrefix = "arXiv",
    primaryClass = "astro-ph.CO",
    doi = "10.1016/j.physletb.2022.137035",
    journal = "Phys. Lett. B",
    volume = "828",
    pages = "137035",
    year = "2022"
}

@article{Ferrante:2022mui,
    author = "Ferrante, Giacomo and Franciolini, Gabriele and Iovino, Junior., Antonio and Urbano, Alfredo",
    title = "{Primordial non-Gaussianity up to all orders: Theoretical aspects and implications for primordial black hole models}",
    eprint = "2211.01728",
    archivePrefix = "arXiv",
    primaryClass = "astro-ph.CO",
    doi = "10.1103/PhysRevD.107.043520",
    journal = "Phys. Rev. D",
    volume = "107",
    number = "4",
    pages = "043520",
    year = "2023"
}

@article{Gow:2022jfb,
    author = "Gow, Andrew D. and Assadullahi, Hooshyar and Jackson, Joseph H. P. and Koyama, Kazuya and Vennin, Vincent and Wands, David",
    title = "{Non-perturbative non-Gaussianity and primordial black holes}",
    eprint = "2211.08348",
    archivePrefix = "arXiv",
    primaryClass = "astro-ph.CO",
    doi = "10.1209/0295-5075/acd417",
    journal = "EPL",
    volume = "142",
    number = "4",
    pages = "49001",
    year = "2023"
}

@article{Pi:2022ysn,
    author = "Pi, Shi and Sasaki, Misao",
    title = "{Logarithmic Duality of the Curvature Perturbation}",
    eprint = "2211.13932",
    archivePrefix = "arXiv",
    primaryClass = "astro-ph.CO",
    reportNumber = "IPMU22-0060, YITP-22-144",
    doi = "10.1103/PhysRevLett.131.011002",
    journal = "Phys. Rev. Lett.",
    volume = "131",
    number = "1",
    pages = "011002",
    year = "2023"
}

@article{Jackson:2022unc,
    author = "Jackson, Joseph H. P. and Assadullahi, Hooshyar and Koyama, Kazuya and Vennin, Vincent and Wands, David",
    title = "{Numerical simulations of stochastic inflation using importance sampling}",
    eprint = "2206.11234",
    archivePrefix = "arXiv",
    primaryClass = "astro-ph.CO",
    doi = "10.1088/1475-7516/2022/10/067",
    journal = "JCAP",
    volume = "10",
    pages = "067",
    year = "2022"
}

@article{Hooshangi:2023kss,
    author = "Hooshangi, Sina and Namjoo, Mohammad Hossein and Noorbala, Mahdiyar",
    title = "{Tail diversity from inflation}",
    eprint = "2305.19257",
    archivePrefix = "arXiv",
    primaryClass = "astro-ph.CO",
    doi = "10.1088/1475-7516/2023/09/023",
    journal = "JCAP",
    volume = "09",
    pages = "023",
    year = "2023"
}

@article{Inui:2024sce,
    author = "Inui, Ryoto and Motohashi, Hayato and Pi, Shi and Tada, Yuichiro and Yokoyama, Shuichiro",
    title = "{Constant roll and non-Gaussian tail in light of logarithmic duality}",
    eprint = "2409.13500",
    archivePrefix = "arXiv",
    primaryClass = "astro-ph.CO",
    doi = "10.1088/1475-7516/2025/02/042",
    journal = "JCAP",
    volume = "02",
    pages = "042",
    year = "2025"
}

@article{Sharma:2024fbr,
    author = "Sharma, Devanshu",
    title = "{Stochastic inflation and non-perturbative power spectrum beyond slow roll}",
    eprint = "2411.08854",
    archivePrefix = "arXiv",
    primaryClass = "astro-ph.CO",
    doi = "10.1088/1475-7516/2025/03/017",
    journal = "JCAP",
    volume = "03",
    pages = "017",
    year = "2025"
}

@article{Miyamoto:2024hin,
    author = "Miyamoto, Koichi and Tada, Yuichiro",
    title = "{Improved quantum algorithm for calculating eigenvalues of differential operators and its application to estimating the decay rate of the perturbation distribution tail in stochastic inflation}",
    eprint = "2410.02276",
    archivePrefix = "arXiv",
    primaryClass = "quant-ph",
    month = "10",
    year = "2024"
}

@article{Animali:2024jiz,
    author = "Animali, Chiara and Vennin, Vincent",
    title = "{Clustering of primordial black holes from quantum diffusion during inflation}",
    eprint = "2402.08642",
    archivePrefix = "arXiv",
    primaryClass = "astro-ph.CO",
    doi = "10.1088/1475-7516/2024/08/026",
    journal = "JCAP",
    volume = "08",
    pages = "026",
    year = "2024"
}

@article{Launay:2024qsm,
    author = "Launay, Yoann L. and Rigopoulos, Gerasimos I. and Shellard, E. Paul S.",
    title = "{Stochastic inflation in general relativity}",
    eprint = "2401.08530",
    archivePrefix = "arXiv",
    primaryClass = "gr-qc",
    doi = "10.1103/PhysRevD.109.123523",
    journal = "Phys. Rev. D",
    volume = "109",
    number = "12",
    pages = "123523",
    year = "2024"
}

@article{Jackson:2024aoo,
    author = "Jackson, Joseph H. P. and Assadullahi, Hooshyar and Gow, Andrew D. and Koyama, Kazuya and Vennin, Vincent and Wands, David",
    title = "{Stochastic inflation beyond slow roll: noise modelling and importance sampling}",
    eprint = "2410.13683",
    archivePrefix = "arXiv",
    primaryClass = "astro-ph.CO",
    doi = "10.1088/1475-7516/2025/04/073",
    journal = "JCAP",
    volume = "04",
    pages = "073",
    year = "2025"
}

@article{Atal:2019cdz,
    author = "Atal, Vicente and Garriga, Jaume and Marcos-Caballero, Airam",
    title = "{Primordial black hole formation with non-Gaussian curvature perturbations}",
    eprint = "1905.13202",
    archivePrefix = "arXiv",
    primaryClass = "astro-ph.CO",
    doi = "10.1088/1475-7516/2019/09/073",
    journal = "JCAP",
    volume = "09",
    pages = "073",
    year = "2019"
}

@article{Atal:2019erb,
    author = "Atal, Vicente and Cid, Judith and Escriv\`a, Albert and Garriga, Jaume",
    title = "{PBH in single field inflation: the effect of shape dispersion and non-Gaussianities}",
    eprint = "1908.11357",
    archivePrefix = "arXiv",
    primaryClass = "astro-ph.CO",
    doi = "10.1088/1475-7516/2020/05/022",
    journal = "JCAP",
    volume = "05",
    pages = "022",
    year = "2020"
}

@article{Animali:2022otk,
    author = "Animali, Chiara and Vennin, Vincent",
    title = "{Primordial black holes from stochastic tunnelling}",
    eprint = "2210.03812",
    archivePrefix = "arXiv",
    primaryClass = "astro-ph.CO",
    doi = "10.1088/1475-7516/2023/02/043",
    journal = "JCAP",
    volume = "02",
    pages = "043",
    year = "2023"
}

@article{Tada:2021zzj,
    author = "Tada, Yuichiro and Vennin, Vincent",
    title = "{Statistics of coarse-grained cosmological fields in stochastic inflation}",
    eprint = "2111.15280",
    archivePrefix = "arXiv",
    primaryClass = "astro-ph.CO",
    doi = "10.1088/1475-7516/2022/02/021",
    journal = "JCAP",
    volume = "02",
    number = "02",
    pages = "021",
    year = "2022"
}

@article{Shibata:1999zs,
    author = "Shibata, Masaru and Sasaki, Misao",
    title = "{Black hole formation in the Friedmann universe: Formulation and computation in numerical relativity}",
    eprint = "gr-qc/9905064",
    archivePrefix = "arXiv",
    reportNumber = "OU-TAP-93",
    doi = "10.1103/PhysRevD.60.084002",
    journal = "Phys. Rev. D",
    volume = "60",
    pages = "084002",
    year = "1999"
}

@article{Harada:2015yda,
    author = "Harada, Tomohiro and Yoo, Chul-Moon and Nakama, Tomohiro and Koga, Yasutaka",
    title = "{Cosmological long-wavelength solutions and primordial black hole formation}",
    eprint = "1503.03934",
    archivePrefix = "arXiv",
    primaryClass = "gr-qc",
    reportNumber = "RUP-15-5, RESCEU-4-15",
    doi = "10.1103/PhysRevD.91.084057",
    journal = "Phys. Rev. D",
    volume = "91",
    number = "8",
    pages = "084057",
    year = "2015"
}

@article{Musco:2018rwt,
    author = "Musco, Ilia",
    title = "{Threshold for primordial black holes: Dependence on the shape of the cosmological perturbations}",
    eprint = "1809.02127",
    archivePrefix = "arXiv",
    primaryClass = "gr-qc",
    doi = "10.1103/PhysRevD.100.123524",
    journal = "Phys. Rev. D",
    volume = "100",
    number = "12",
    pages = "123524",
    year = "2019"
}

@article{Raatikainen:2023bzk,
    author = {Raatikainen, Sami and R\"as\"anen, Syksy and Tomberg, Eemeli},
    title = "{Primordial Black Hole Compaction Function from Stochastic Fluctuations in Ultraslow-Roll Inflation}",
    eprint = "2312.12911",
    archivePrefix = "arXiv",
    primaryClass = "astro-ph.CO",
    reportNumber = "HIP-2023-18/TH",
    doi = "10.1103/PhysRevLett.133.121403",
    journal = "Phys. Rev. Lett.",
    volume = "133",
    number = "12",
    pages = "121403",
    year = "2024"
}

@article{Noorbala:2018zlv,
    author = "Noorbala, Mahdiyar and Vennin, Vincent and Assadullahi, Hooshyar and Firouzjahi, Hassan and Wands, David",
    title = "{Tunneling in Stochastic Inflation}",
    eprint = "1806.09634",
    archivePrefix = "arXiv",
    primaryClass = "hep-th",
    doi = "10.1088/1475-7516/2018/09/032",
    journal = "JCAP",
    volume = "09",
    pages = "032",
    year = "2018"
}

@article{Vennin:2015hra,
    author = "Vennin, Vincent and Starobinsky, Alexei A.",
    title = "{Correlation Functions in Stochastic Inflation}",
    eprint = "1506.04732",
    archivePrefix = "arXiv",
    primaryClass = "hep-th",
    doi = "10.1140/epjc/s10052-015-3643-y",
    journal = "Eur. Phys. J. C",
    volume = "75",
    pages = "413",
    year = "2015"
}

@article{Helmer:2006tz,
    author = "Helmer, Ferdinand and Winitzki, Sergei",
    title = "{Self-reproduction in k-inflation}",
    eprint = "gr-qc/0608019",
    archivePrefix = "arXiv",
    doi = "10.1103/PhysRevD.74.063528",
    journal = "Phys. Rev. D",
    volume = "74",
    pages = "063528",
    year = "2006"
}

@article{Winitzki:1995pg,
    author = "Winitzki, Serge and Vilenkin, Alexander",
    title = "{Uncertainties of predictions in models of eternal inflation}",
    eprint = "gr-qc/9510054",
    archivePrefix = "arXiv",
    reportNumber = "TUTP-95-03",
    doi = "10.1103/PhysRevD.53.4298",
    journal = "Phys. Rev. D",
    volume = "53",
    pages = "4298--4310",
    year = "1996"
}

@article{Nambu:1988je,
    author = "Nambu, Yasusada and Sasaki, Misao",
    title = "{Stochastic approach to chaotic inflation and the distribution of universes}",
    reportNumber = "RRK-88-35",
    doi = "10.1016/0370-2693(89)90385-7",
    journal = "Phys. Lett. B",
    volume = "219",
    pages = "240--246",
    year = "1989"
}

@article{Nambu:1989uf,
    author = "Nambu, Yasusada",
    title = "{Stochastic Dynamics of an Inflationary Model and Initial Distribution of Universes}",
    reportNumber = "RRK-89-3",
    doi = "10.1143/PTP.81.1037",
    journal = "Prog. Theor. Phys.",
    volume = "81",
    pages = "1037",
    year = "1989"
}

@book{Birrell:1982ix,
    author = "Birrell, N. D. and Davies, P. C. W.",
    title = "{Quantum Fields in Curved Space}",
    doi = "10.1017/CBO9780511622632",
    isbn = "978-0-511-62263-2, 978-0-521-27858-4",
    publisher = "Cambridge University Press",
    address = "Cambridge, UK",
    series = "Cambridge Monographs on Mathematical Physics",
    year = "1982"
}

@article{Boddy:2016zkn,
    author = "Boddy, Kimberly K. and Carroll, Sean M. and Pollack, Jason",
    title = "{How Decoherence Affects the Probability of Slow-Roll Eternal Inflation}",
    eprint = "1612.04894",
    archivePrefix = "arXiv",
    primaryClass = "hep-th",
    reportNumber = "CALT-2016-033, UH-511-1272-2016",
    doi = "10.1103/PhysRevD.96.023539",
    journal = "Phys. Rev. D",
    volume = "96",
    number = "2",
    pages = "023539",
    year = "2017"
}

@article{Inui:2024fgk,
    author = "Inui, Ryoto and Joana, Cristian and Motohashi, Hayato and Pi, Shi and Tada, Yuichiro and Yokoyama, Shuichiro",
    title = "{Primordial black holes and induced gravitational waves from logarithmic non-Gaussianity}",
    eprint = "2411.07647",
    archivePrefix = "arXiv",
    primaryClass = "astro-ph.CO",
    doi = "10.1088/1475-7516/2025/03/021",
    journal = "JCAP",
    volume = "03",
    pages = "021",
    year = "2025"
}

@book{arfken2011mathematical,
  title={Mathematical methods for physicists: a comprehensive guide},
  author={Arfken, George B and Weber, Hans J and Harris, Frank E},
  year={2011},
  publisher={Academic press}
}

@book{sturmliouville,
  title={Sturm--Liouville Theory and its Applications},
  author={Al-Gwaiz, M A},
  year={2008},
  publisher={Springer London}
}

@article{Hawking:1981fz,
    author = "Hawking, S. W. and Moss, I. G.",
    title = "{Supercooled Phase Transitions in the Very Early Universe}",
    reportNumber = "Print-82-0181 (CAMBRIDGE)",
    doi = "10.1016/0370-2693(82)90946-7",
    journal = "Phys. Lett. B",
    volume = "110",
    pages = "35--38",
    year = "1982"
}

@article{Boubekeur:2005zm,
    author = "Boubekeur, Lotfi and Lyth, David. H.",
    title = "{Hilltop inflation}",
    eprint = "hep-ph/0502047",
    archivePrefix = "arXiv",
    doi = "10.1088/1475-7516/2005/07/010",
    journal = "JCAP",
    volume = "07",
    pages = "010",
    year = "2005"
}

@article{Rigopoulos:2021nhv,
    author = "Rigopoulos, Gerasimos and Wilkins, Ashley",
    title = "{Inflation is always semi-classical: diffusion domination overproduces Primordial Black Holes}",
    eprint = "2107.05317",
    archivePrefix = "arXiv",
    primaryClass = "astro-ph.CO",
    doi = "10.1088/1475-7516/2021/12/027",
    journal = "JCAP",
    volume = "12",
    number = "12",
    pages = "027",
    year = "2021"
}

@article{Blachier:2025tcq,
    author = "Blachier, Baptiste and Ringeval, Christophe",
    title = "{Time-reversed Stochastic Inflation}",
    eprint = "2504.17680",
    archivePrefix = "arXiv",
    primaryClass = "astro-ph.CO",
    month = "4",
    year = "2025"
}

\end{document}